\documentclass[a4paper]{jpconf}
\usepackage[dvipdfmx]{graphicx}
\usepackage{bm}
\usepackage{braket}
\usepackage{fancyhdr}
\usepackage{amsmath}
\usepackage{cases}
\usepackage{color}

\begin{document}

\title{Edge states of mechanical diamond and its topological origin}

\author{Yuta Takahashi$^1$, Toshikaze Kariyado$^2$, and Yasuhiro Hatsugai$^1$}

\address{$^1$Graduate School of Pure and Applied Science, University of Tsukuba, Tsukuba 305-8571, Japan\\
}
\address{$^2$International Center for Materials Nanoarchitectonics (WPI-MANA), National Institute for Materials Science, Tsukuba 305-0044, Japan\\
}

\ead{takahashi@rhodia.ph.tsukuba.ac.jp}

\begin{abstract}
  A mechanical diamond,  a classical mechanics of
  a spring-mass model arrayed on a diamond lattice, is
  discussed topologically.
  Its frequency dispersion possesses an intrinsic nodal structure in the three-dimensional Brillouin zone (BZ)
  protected by  the chiral symmetry. Topological changes of the line nodes are
  demonstrated associated with modification of the tension.
  The line nodes projected into two-dimensional BZ form loops which
  are characterized by  the quantized Zak phases by 0 and $\pi$.
  With boundaries,  
  edge states are discussed in relation to the Zak phases and winding numbers.
  It establishes a bulk-edge correspondence of the mechanical diamond.
  \end{abstract}

\section{Introduction}
Topological semimetal is a system in which its band gap is finite almost
everywhere in the Brillouin zone, except on some sets of isolated
points. In short, it is a system with singular gapless points
\cite{Wallace1947}. Generally, existence of a singular gapless point
leads to nontrivial topology. That is, a singularity often serves as a
source of a ``twist'' of Bloch wave functions captured by the Berry
curvature, which gives rise to nontrivial
topology.  When a system is
characterized by a point-like singularity associated with linear
dispersion, i.e., a Dirac cone, it is Dirac/Weyl semimetal, which is a
representative topological semimetal
\cite{Young2012,Wang2012,Liu2014,BurkovWeyl2011,Wan2011}. Very recently,
the other kinds of the topological semimetal, nodal line semimetal where
the gapless points form a line, typically a closed loop, begins to attract
attention as a new stage to play with topology
\cite{Burkovnodal2011}. Just as in the case of fully
gapped topological insulators, topological semimetals are characterized
by topologically protected edge modes as a consequence of the bulk-edge
correspondence \cite{YH93}.

On the other hand, there are  intensive efforts to
export the idea of topological insulators and 
topological semimetals to classical world \cite{PhysRevLett.100.013904,PhysRevLett.100.013905,Prodan2009,Berg2011,Kane2014}. Especially, classical
mechanical systems are interesting playgrounds because of its simplicity
and flexibility to tune parameters
\cite{Prodan2009,Berg2011,Kane2014,Chen2014,WangYao2015,WangPai2015,Susstrunk2015,Nash2015,Paulose2015,Chen2016,Po2016}.
In fact, we have already seen topological edge modes in
various mechanical systems
\cite{Prodan2009,Berg2011,Kane2014,WangYao2015,WangPai2015,Susstrunk2015,Nash2015}.
One specific example to show the easy-to-tune feature is
\textit{mechanical graphene}, which is a honeycomb
spring-mass model
\cite{Cserti2004,WangPai2015,Nash2015,Kariyado:2015aa,Socolar2016}. In
mechanical graphene, Dirac cones are known to exist, and their number
and positions in the Brillouin zone can be
controlled by simply changing the tension of springs in equilibrium
\cite{Kariyado:2015aa}. It is worth noting that the
effect of the equilibrium tension can be interpreted as a spin-orbit
coupling if clockwise and counterclockwise motion of the mass point is
mapped to spin \cite{1367-2630-18-11-113014,Salerno2016}.

In this paper, we make an analysis on a spring-mass model with the
diamond structure, namely, \textit{mechanical diamond}, which is a
natural extension of mechanical graphene to three-dimension \cite{Kariyado:2015aa}. 
Mechanical diamond is a typical classical counterpart of a nodal line semimetal. It
is revealed that the equilibrium tension induces
unique evolution of the structure of the gapless line nodes. Topological
properties of mechanical diamond are also investigated by relating edge
modes, the quantized Zak phase, and the winding number. In accordance
with the unique line node structure, the edge modes also shows unique
distribution on the surface Brillouin zone, including the situation with
the edge mode multiplicity 2 or 3. We confirm that these features are
well captured by the quantized Zak phase and the winding number, which
establishes the bulk-edge correspondence in mechanical diamond.

The paper is organized as follows. In Sec.~\ref{sec:formulation}, basic
notions of mechanical diamond are explained. Then, the frequency
dispersion is shown in
Sec.~\ref{sec:dispersion}. Section~\ref{sec:topological} is devoted to
the topological arguments, and the paper is summarized in Sec.~\ref{sec:summary}.

\section{Formulation}\label{sec:formulation}

\begin{figure}[htbp]
 \begin{minipage}{.4\hsize}
 \centering
    \includegraphics[clip,scale=0.28]{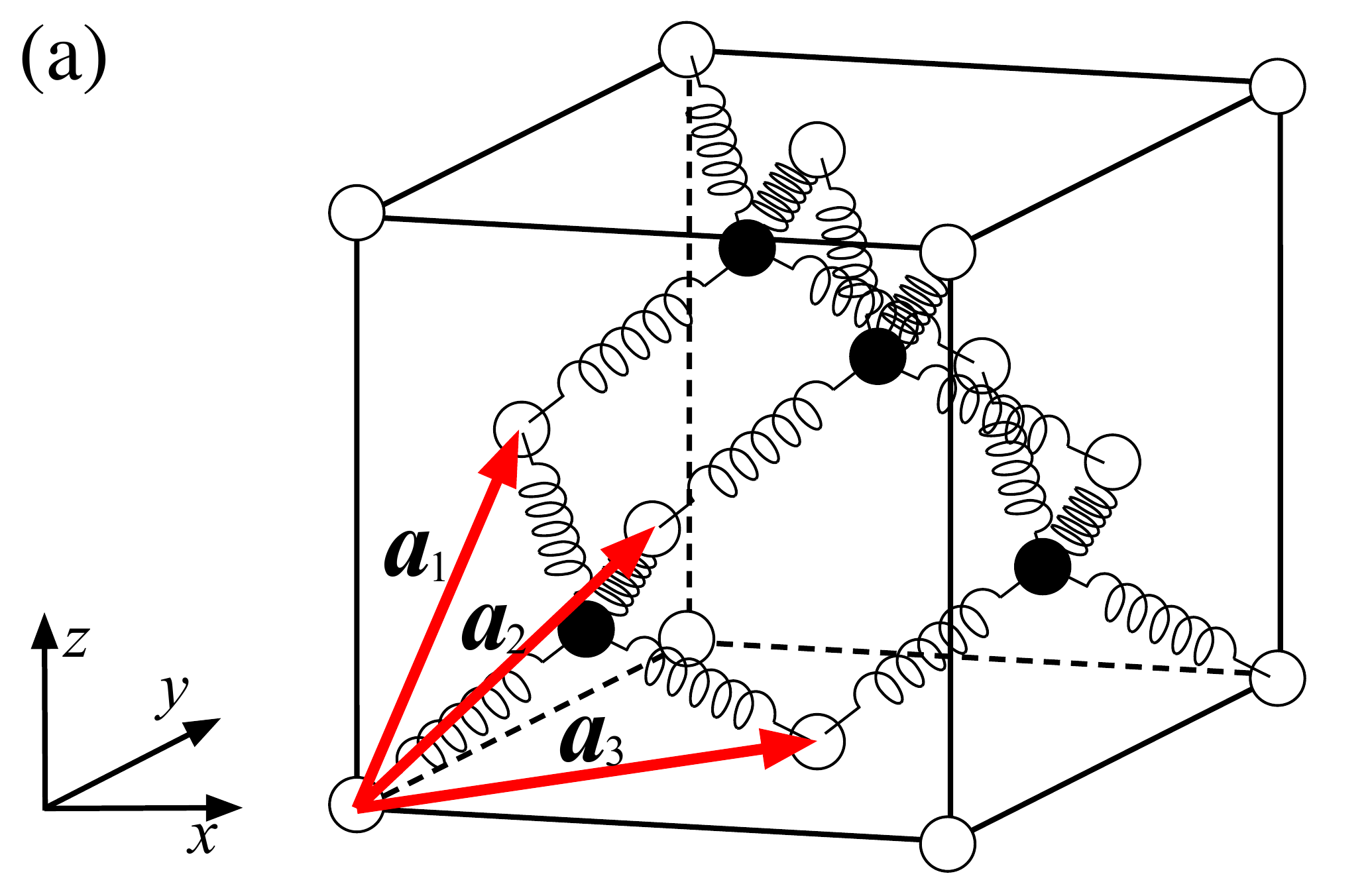}
    \end{minipage}
 \begin{minipage}{.3\hsize}
  \centering
   \includegraphics[clip,scale=0.21]{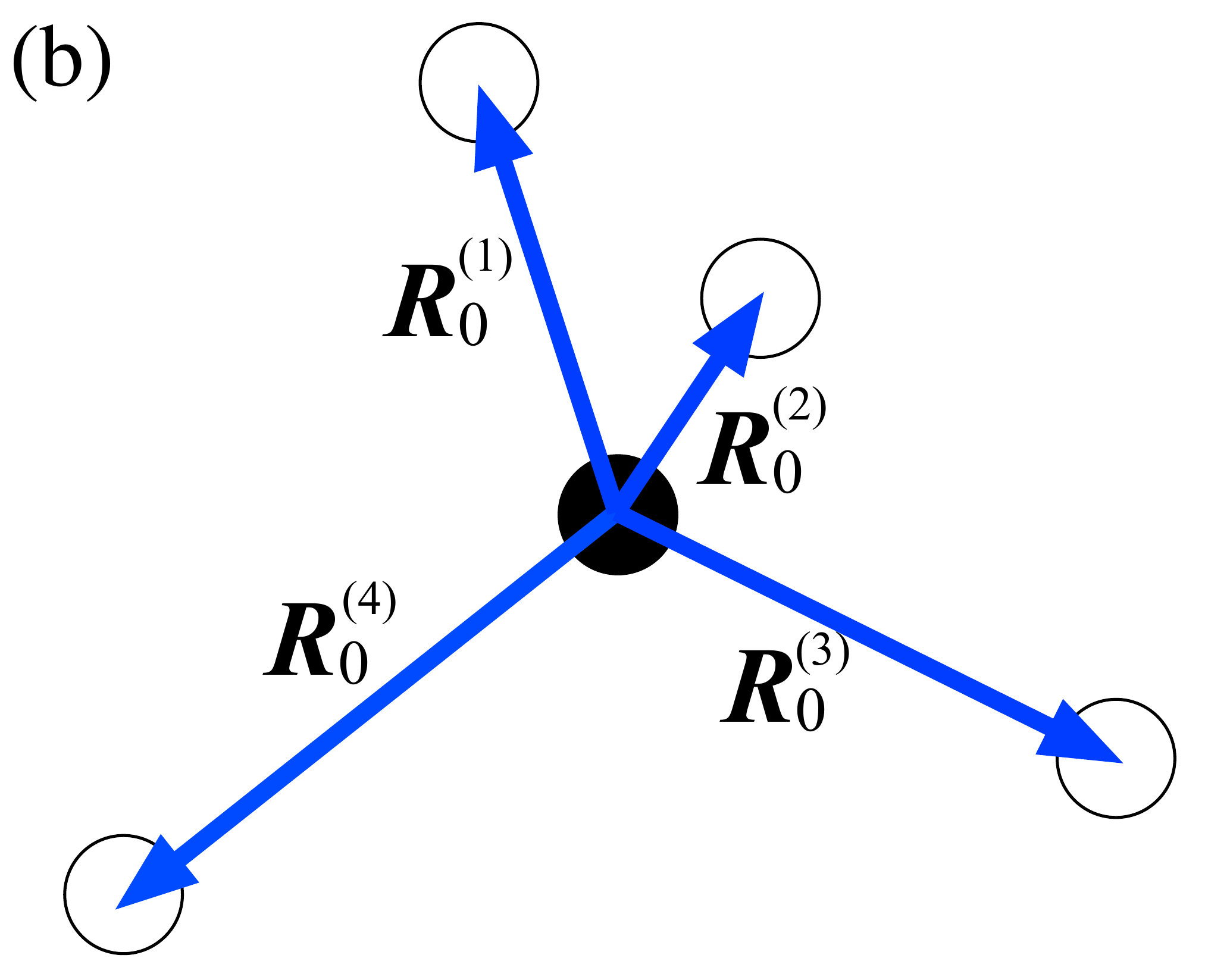}
 \end{minipage}
 \begin{minipage}{.3\hsize}
 \centering
    \includegraphics[clip,scale=0.13]{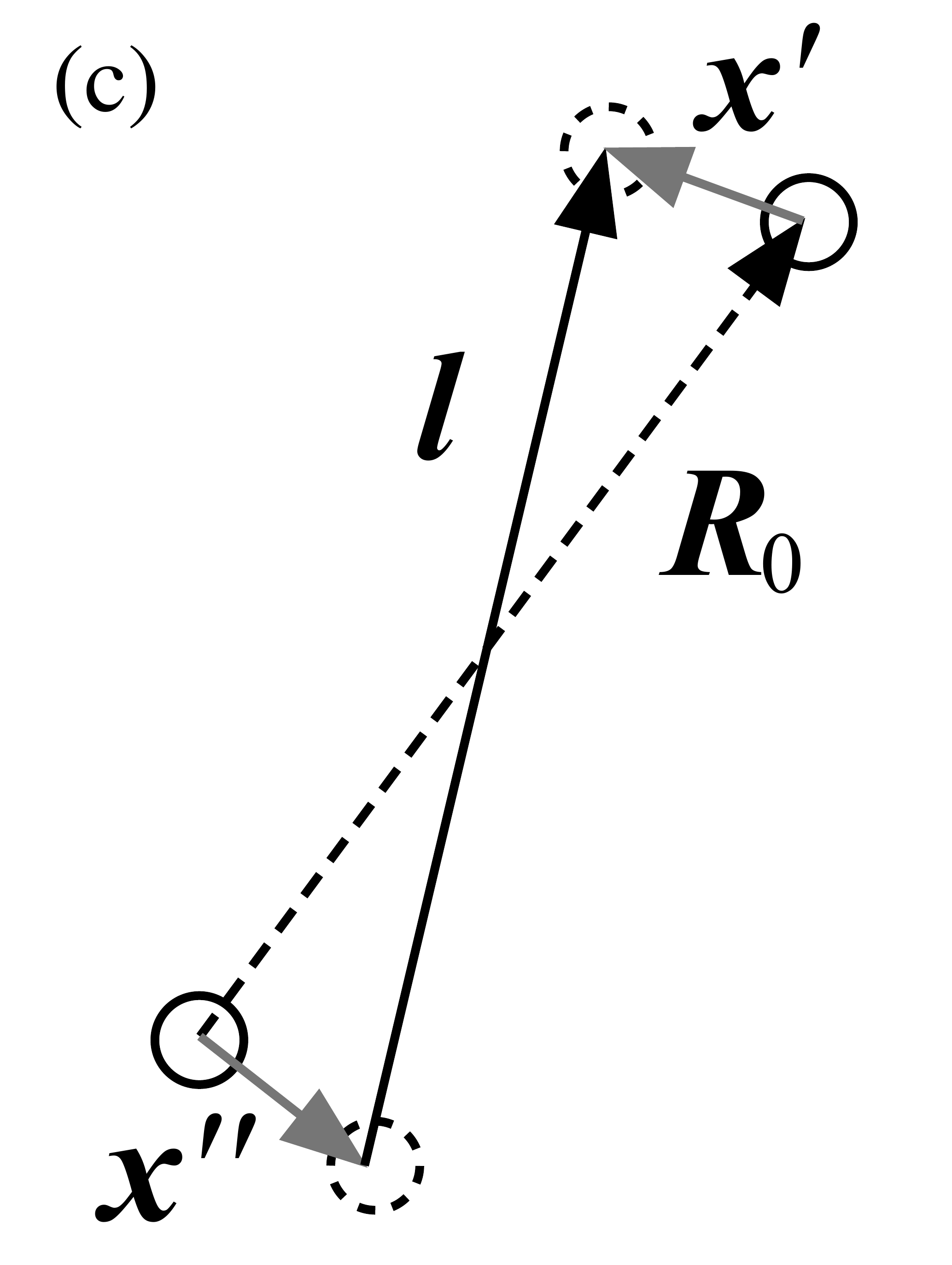}
 \end{minipage}
 \caption{
 (a) Schematic picture for mechanical diamond. The unit
 translation vectors are also shown as $\bm{a}_i$. (b) Nearest neighbor
 vectors $\bm{R}^{(i)}_0$. (c) Definitions of $\bm R_0$, $\bm x'$ and
 $\bm x''$.}
    \label{fig:setup}
\end{figure}

Let us introduce our model, \textit{mechanical diamond},
which consists of mass points aligned in the diamond structure and
springs connecting the nearest neighbor pairs of the mass points [See
Fig.~\ref{fig:setup}(a)].  This
three-dimensional model is a natural extension of two-dimensional
\textit{mechanical graphene} \cite{Kariyado:2015aa}, a spring-mass model
with honeycomb structure. As in the case of 2D mechanical graphene,
parameters characterizing our model are mass of mass points $m$
($m$ is fixed to unit for simplicity), spring
constant $\kappa$, natural length of springs $l_0$ and distance between
the neighboring mass points $R_0$. Note that $R_0$ and $l_0$ does not
necessarily match with each other, namely, $R_0$ can be larger than
$l_0$ if we apply a proper boundary condition to exert uniform
outward tension to the system. In order to investigate the dynamics of
the system, we introduce a quantity
$\bm{x}_{\bm{R}a}={}^t\!(x_{\bm{R}a},y_{\bm{R}a},z_{\bm{R}a})$
describing displacement of the each mass point from the equilibrium
position. Here, $\bm{R}$ designates a lattice point and $a$ is a
sublattice index.

We assume that the elastic energy $U_s$ of a specific spring can be
expressed as $U_s=\frac{1}{2}\kappa(l-l_0)^2$ with 
$l=|\bm{R}_0+\bm{x}'-\bm{x}''|$, where $\bm{x}'$, $\bm{x}''$, and
$\bm{R}_0$ are two displacement vectors for two mass points, and a
vector connecting the two equilibrium positions, respectively. By
expanding $U_s$ up to the second order in
$\delta\bm{x}=\bm{x}'-\bm{x}''$, we obtain
\begin{equation}
  U_s \simeq \frac{1}{2}\kappa((R_0-l_0)^2+2(R_0-l_0)\hat{\bm R}_0 \cdot \delta \bm x+\delta x_\mu \gamma_{\bm R_0}^{\mu \nu}\delta x_{\nu}),
  \label{eq:Us}
\end{equation}
where $\gamma_{\bm R_0}^{\mu \nu}=(1-\eta)\delta^{\mu \nu}+\eta
\hat{R}_0^{\mu} \hat{R}_0^{\nu}$, $\hat{\bm R}_0=\bm R_0/|\bm R_0|$ and
$\eta \equiv l_0/R_0$. Here, we implicitly take the summation over $\mu$
and $\nu$, which run through the directions of displacement $x$, $y$,
and $z$, i.e., we apply the Einstein summation convention. The second
term that is linear in $\delta\bm{x}$ has no contribution to the
Newton's equation of motion as far as we focus on displacement from the
equilibrium point, since the linear terms cancel out with each other in
the equation of
motion. The third term is characterized by the parameter $\eta$. For
$\eta=1$, at which the springs have the natural length in equilibrium,
the angle between $\bm{R}_0$ and $\delta\bm{x}$ has significant impact
on the elastic energy because of the factor
$\hat{R}^\mu_0\hat{R}^\nu_0$. On the other hand for $\eta=0$, the
limiting case of the stretched springs in equilibrium, $U_s$ has no
dependence on the direction of $\delta\bm{x}$, in contrast to the case
of $\eta=1$. Therefore, the parameter $\eta$ enables us to tune the
property of the spring-mass model. 


For mechanical diamond, the equation of motion can be derived by
evaluating the elastic energy substituting $\bm{R}^{(i)}_0$ in
Fig.~\ref{fig:setup}(b) into Eq.~\eqref{eq:Us}. If we further assume
that the system is periodic in space and time, namely, by introducing
$\phi_{a\mu}(\bm{k})$ as
$u_{\bm{k} a}^{\mu}=\mathrm{e}^{\mathrm{i}\omega{t}}\phi_{a\mu}(\bm{k})$
and
$x_{\bm{R}a}^{\mu}=\frac{1}{N}\sum_{\bm{k}}\mathrm{e}^{\mathrm{i}\bm{k}\cdot\bm{R}}u_{\bm{k}a}^{\mu}$,
the equation of motion reduces to
\begin{equation}
 -\omega^2\phi_{a\mu}(\bm{k})+\sum_{b}\Gamma_{ab}^{\mu\nu}(\bm{k})\phi_{b\nu}(\bm{k})=0,
\end{equation}
where $\Gamma_{ab}^{\mu\nu}(\bm{k})=(\hat{\Gamma}_{\bm{k}})_{a\mu;b\nu}$,
\begin{equation}
\hat{\Gamma}(\bm{k})=\kappa(4-\frac{8}{3}\eta)\hat{1}+
\begin{pmatrix}
\hat{0} &\hat{\Gamma}_{AB}(\bm{k})\\
\hat{\Gamma}_{AB}^{\dagger} (\bm{k})&\hat{0}
\end{pmatrix},
\label{eq:Gamma_all}
\end{equation}
and
$\hat{\Gamma}_{AB}(\bm{k})=-\kappa(\hat{\gamma}_4+\mathrm{e}^{-\mathrm{i}\bm{k}\cdot\bm{a}_1}\hat{\gamma}_1+\mathrm{e}^{-\mathrm{i}\bm{k}\cdot\bm{a}_2}\hat{\gamma}_2+\mathrm{e}^{-\mathrm{i}\bm{k}\cdot\bm{a}_3}\hat{\gamma}_3)$,
with $\hat{\gamma}_i\equiv\hat{\gamma}_{\bm{R}_0^{(i)}}$ $(i=1,2,3,4)$.
[See Figs.~\ref{fig:setup}(a) and \ref{fig:setup}(b).] The
eigenfrequency and the eigenmodes are calculated by diagonalizing
$\hat{\Gamma}(\bm{k})$. Note that
$\hat{\Gamma}'(\bm{k})=\hat{\Gamma}(\bm{k})-\kappa(4-\frac{8}{3}\eta)\hat{1}$
anticommutes with $\hat{\Upsilon}=\mathrm{diag}(1,1,1,-1,-1,-1)$, i.e.,
$\hat{\Gamma}'(\bm{k})$ has the chiral symmetry. This chiral symmetry
can be used to make topological characterization of
$\hat{\Gamma}(\bm{k})$, since the shift proportional to $\hat{1}$ does
not modify the eigenmodes.

\section{Frequency dispersion and nodal line}\label{sec:dispersion}
 \begin{figure}[tbp]
  \centering
   \includegraphics[clip,scale=0.24]{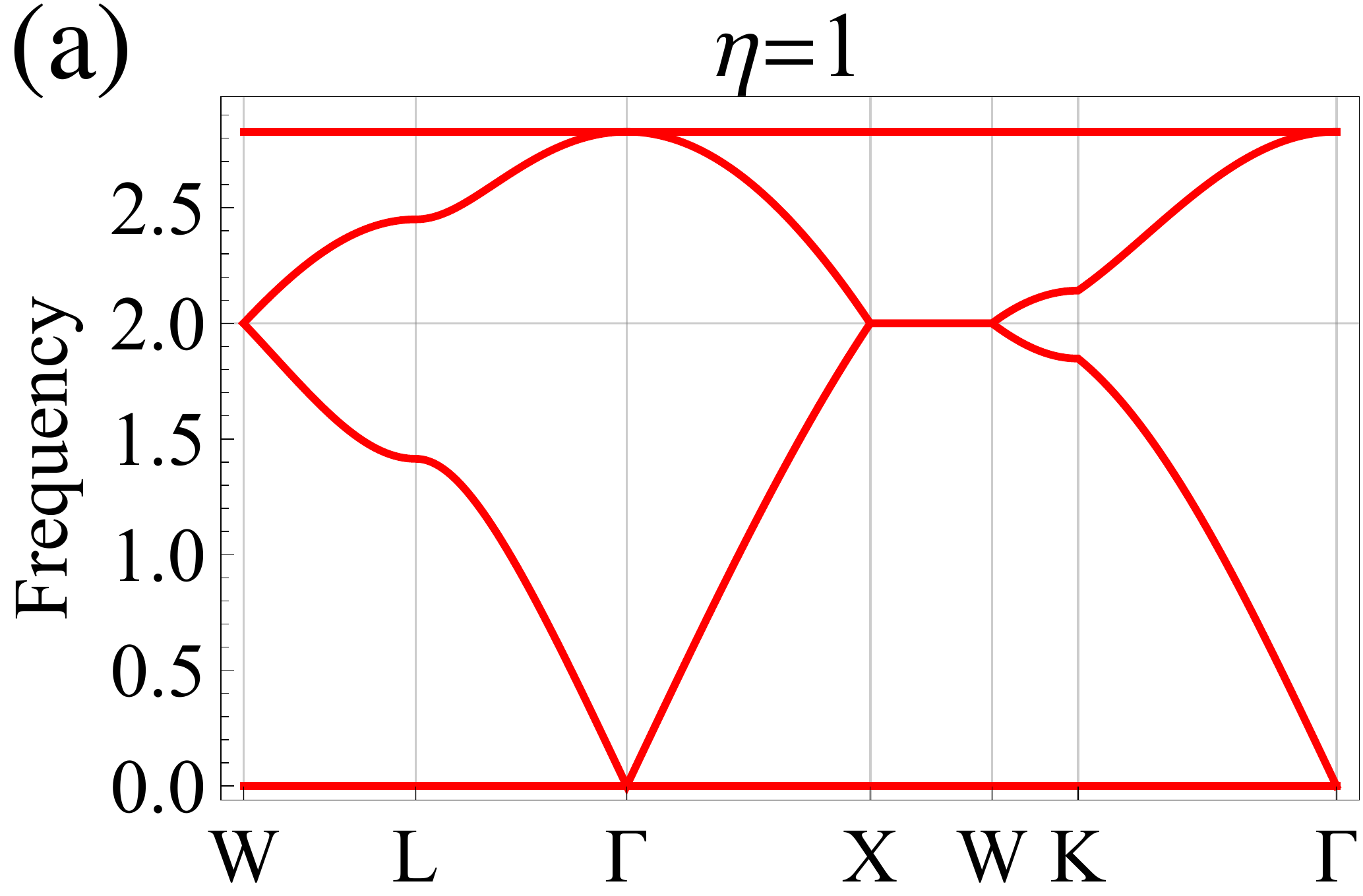}
   \includegraphics[clip,scale=0.24]{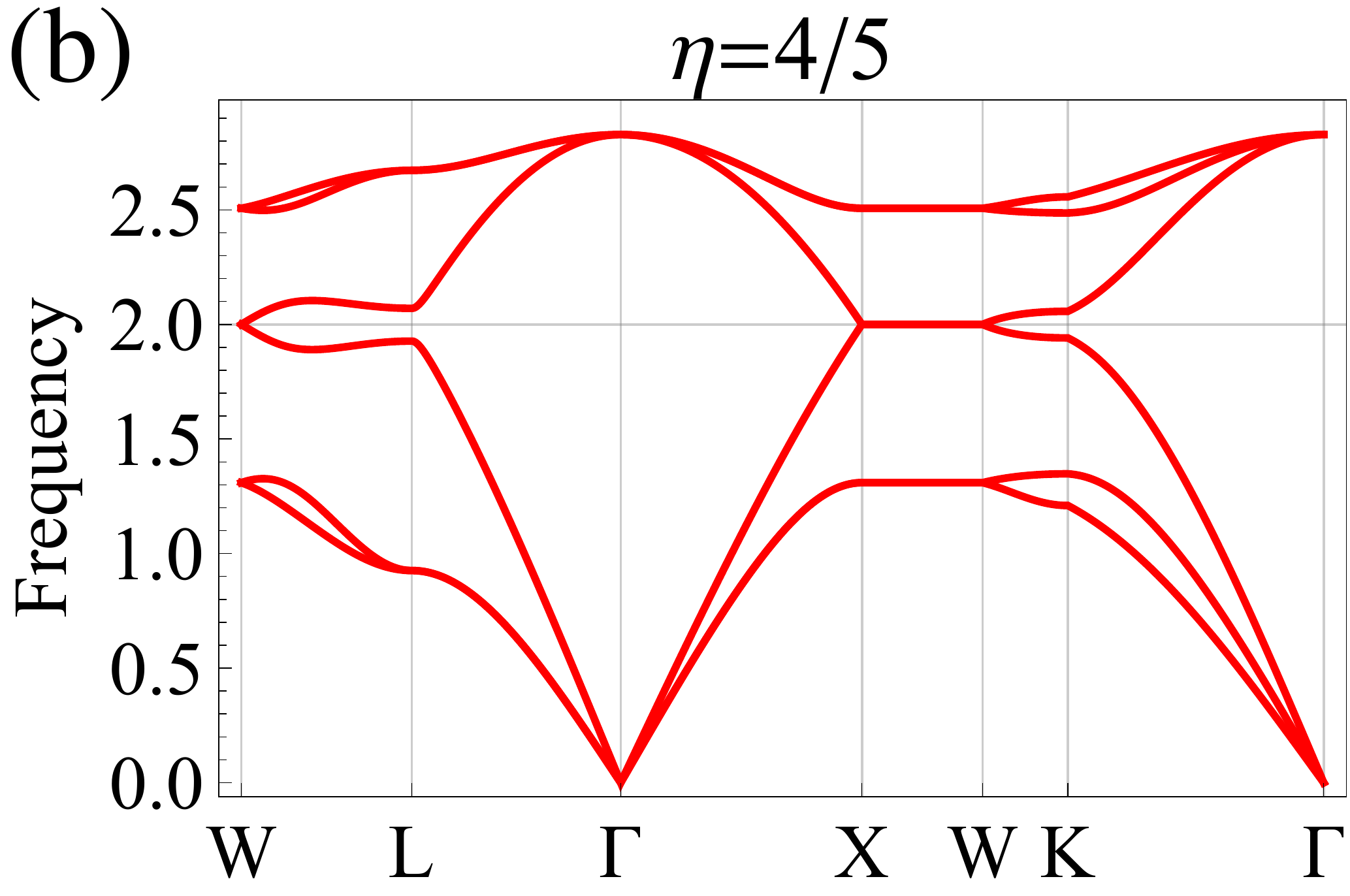}
   \includegraphics[clip,scale=0.24]{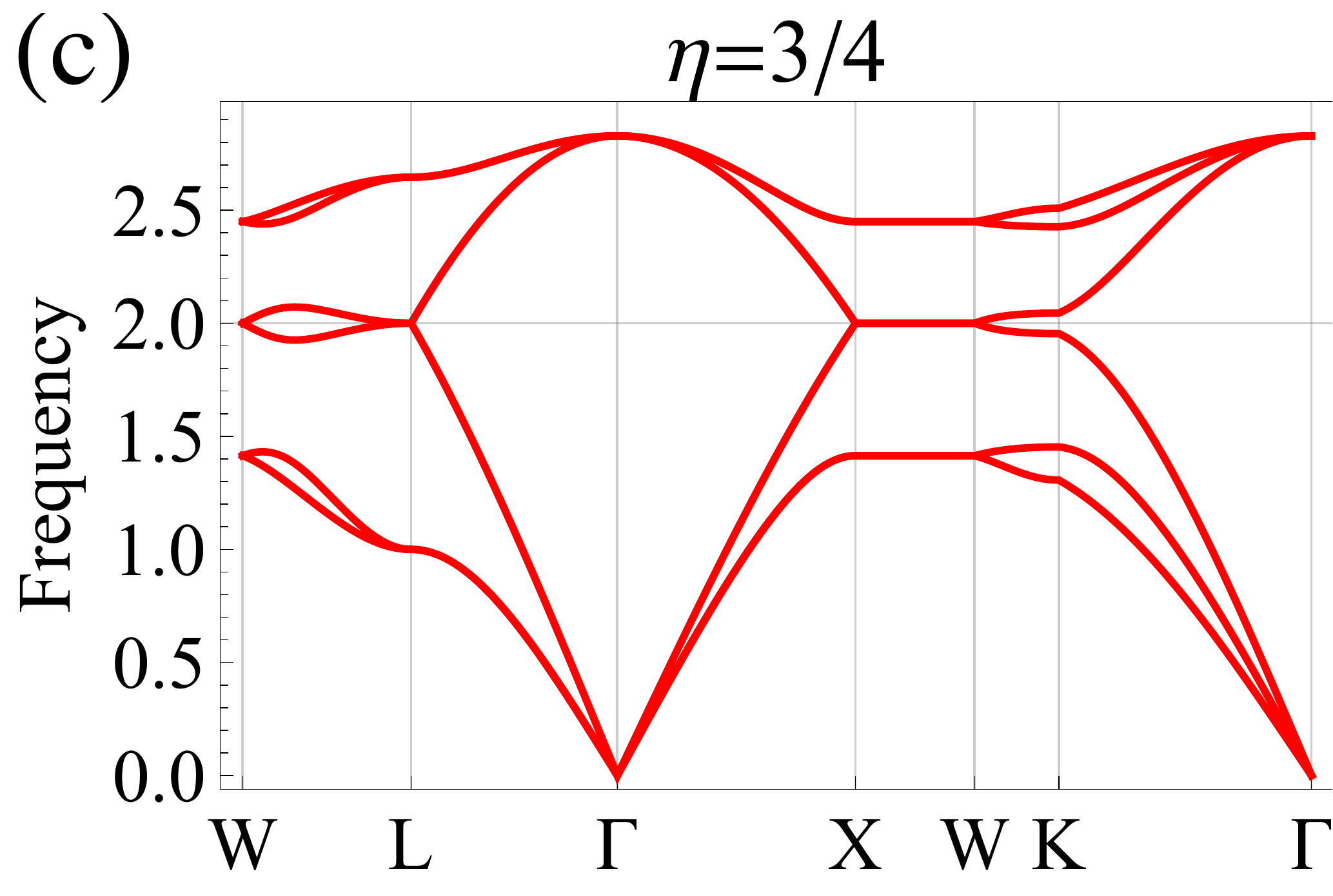}\\
   \includegraphics[clip,scale=0.24]{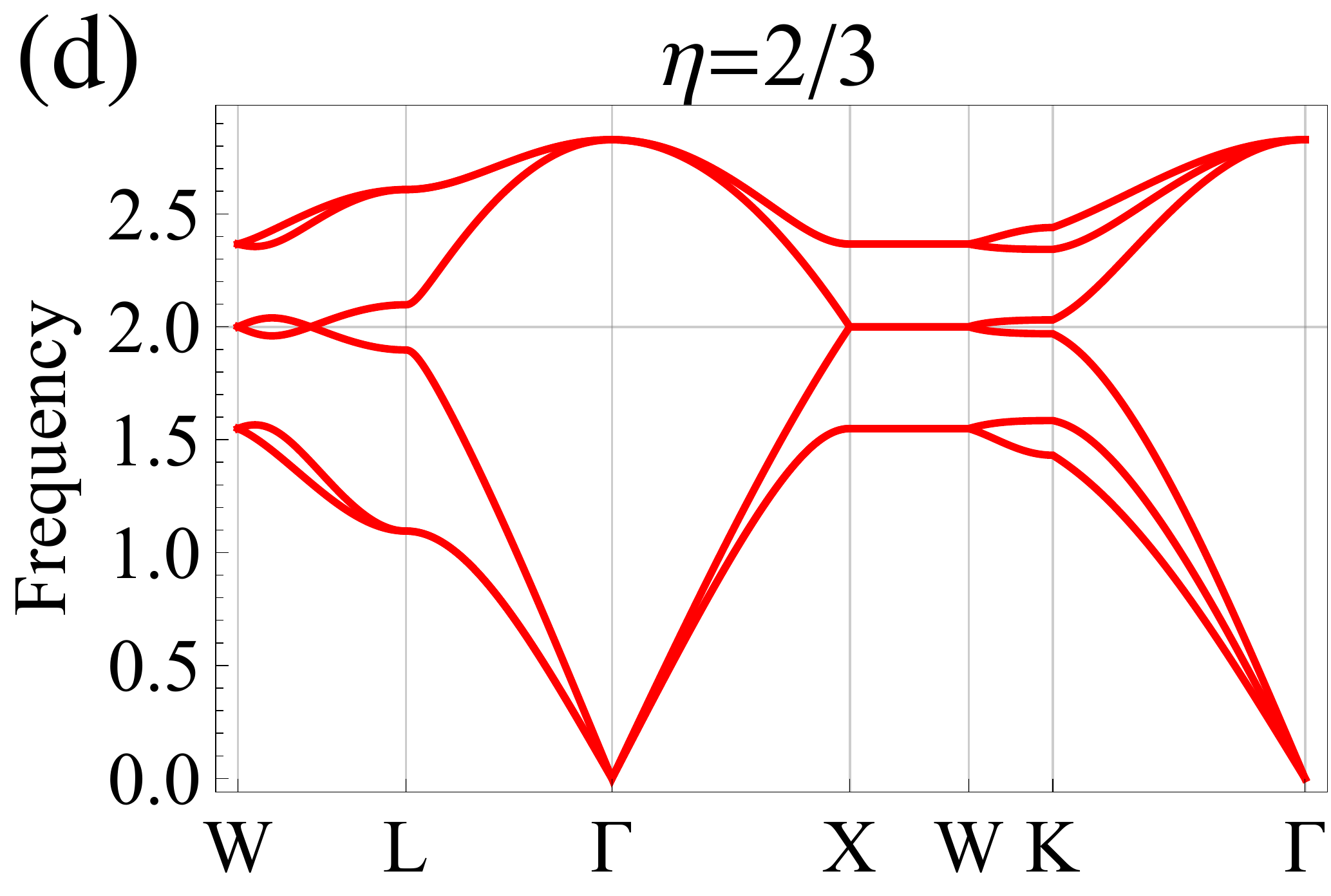}
   \includegraphics[clip,scale=0.24]{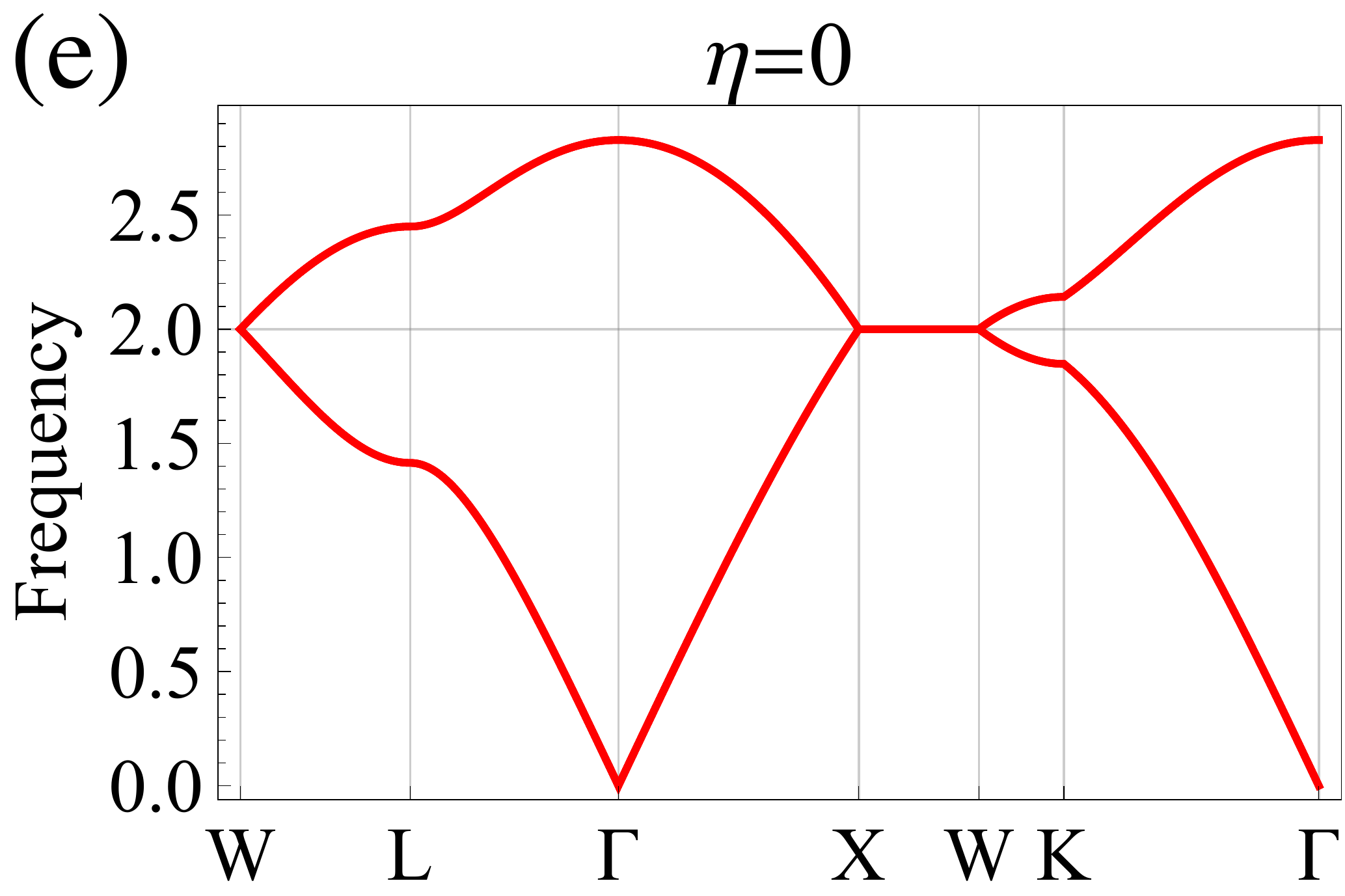}
  \caption{Band structure for (a) $\eta=1$, (b) $\eta=\frac{4}{5}$, (c) $\eta=\frac{3}{4}$, (d) $\eta=\frac{2}{3}$ and (e) $\eta=0$.}
\label{fig:dispersion}
\end{figure}


Figure~\ref{fig:dispersion} shows the frequency band dispersion of the
system for several values of $\eta$. Here, $\kappa$ is scaled by a
factor $1/(1-\frac{2}{3}\eta)$ to remove the $\eta$ dependence of the
total band width. Since we have six degrees of freedom, two from the
sublattices and three from the directions $x$, $y$, and $z$, per a unit
cell, we find six bands. As in the case of the single orbital
tight-binding model with the diamond structure, the gap between the
third and fourth bands is closed on the X-W line regardless of the value
of $\eta$. (We employ the standard notation for the high symmetry points
in the fcc Brillouin zone.) Around these gapless points, the gap grows
linearly in the direction perpendicular to the X-W line. Actually, in
the 3D Brillouin zone, the gapless points form line nodes, which are
protected by the chiral symmetry. This point will be discussed in detail
soon later. Interestingly, a new gapless point, which again forms a line
node, is identified on the W-L line for $\eta<3/4$, namely, we can
generate a line node that is absent in the single orbital tight-binding
model on the diamond lattice simply by applying tension to control
$\eta$.


In order to obtain a global mapping of the geometry of the line nodes in
the 3D Brillouin zone, we employ the Berry phase. This is because the
direct assessment of degeneracy in the eigenfrequency spectrum requires
some care, and it is safer to detect a ``twist'' in the eigenmodes
associated with the singular degeneracy, which is captured by the Berry
phase. In the numerical calculation of the Berry phase, we follow the
idea in Ref.~\cite{doi:10.1143/JPSJ.74.1674}. In the following, we
describe the procedure to map the line nodes in order.


First, the Brillouin zone is decomposed into small cubes whose corners are
specified by $\bm{k}_{\bm{l}}=(k_{l_1},k_{l_2},k_{l_3})$ where
$\bm{l}=(l_1,l_2,l_3)$, $k_{l_\mu}=2\pi{l}_\mu/N_B$, and
$l_{\mu}=0,...,N_{B}-1$. We also introduce a triplet
$\psi(\bm{k})=(\ket{n_1(\bm{k})},\ket{n_2(\bm{k})},\ket{n_3(\bm{k})})$,
where $|n_i(\bm{k})\rangle$ is the eigenvector of the $i$th band at
$\bm{k}$. Here, ``triplet'' because we are focusing on the gap
between the third and fourth bands. Then, we define a $U(1)$ link
variable by
\begin{equation}
 U_{\pm\hat{e}_\mu}(\bm{k})\equiv
  \frac{1}{\mathcal{N}_{\pm\hat{e}_\mu}(\bm{k})}
  \det\psi^{\dagger}(\bm{k})\psi(\bm{k}\pm\frac{\hat{e}_\mu}{N_D}).
 \label{eq:link}
\end{equation}
Here, we have introduced a shorthand notation
$\hat{e}_\mu=(2\pi/N_B)(\delta_{1\mu},\delta_{2\mu},\delta_{3\mu})$, and
$\mathcal{N}_{\pm\hat{e}_\mu}(\bm{k})\equiv|\det\psi^{\dagger}(\bm{k})\psi(\bm{k}\pm\frac{\hat{e}_\mu}{N_D})|$.
Now, we can assign the Berry phase to the each square surface of a small
cube by taking the edge of the square as an integration path to define
the Berry phase. For instance, for the square spanned by
$\bm{k}_{\bm{l}}$, $\bm{k}_{\bm{l}}+\hat{e}_\mu$,
$\bm{k}_{\bm{l}}+\hat{e}_\nu$, and
$\bm{k}_{\bm{l}}+\hat{e}_\mu+\hat{e}_\nu$ ($\mu\neq\nu$), the assigned
Berry phase is computed as 
\begin{multline}
 \tilde{\upsilon}_{\mu\nu}(\bm{k}_{\bm{l}})
 \equiv \sum_{a=0}^{N_D-1}\mathrm{Arg}\, U_{\hat{e}_\mu}(\bm{k}_{\bm{l}}+\frac{a}{N_D}\hat{e}_\mu)
 +\sum_{a=0}^{N_D-1}\mathrm{Arg}\, U_{\hat{e}_\nu}(\bm{k}_{\bm{l}}+\hat{e}_\mu+\frac{a}{N_D}\hat{e}_\nu)\\
 +\sum_{a=0}^{N_D-1}\mathrm{Arg}\,
 U_{-\hat{e}_\mu}(\bm{k}_{\bm{l}}+\hat{e}_\mu+\hat{e}_\nu-\frac{a}{N_D}\hat{e}_\mu)
 +\sum_{a=0}^{N_D-1}\mathrm{Arg}\, U_{-\hat{e}_\nu}(\bm{k}_{\bm{l}}+\hat{e}_\nu-\frac{a}{N_D}\hat{e}_\nu).
\label{eq:dis_bp}
\end{multline}

 \begin{figure}[htbp]
  \centering
  \includegraphics[clip,scale=0.4]{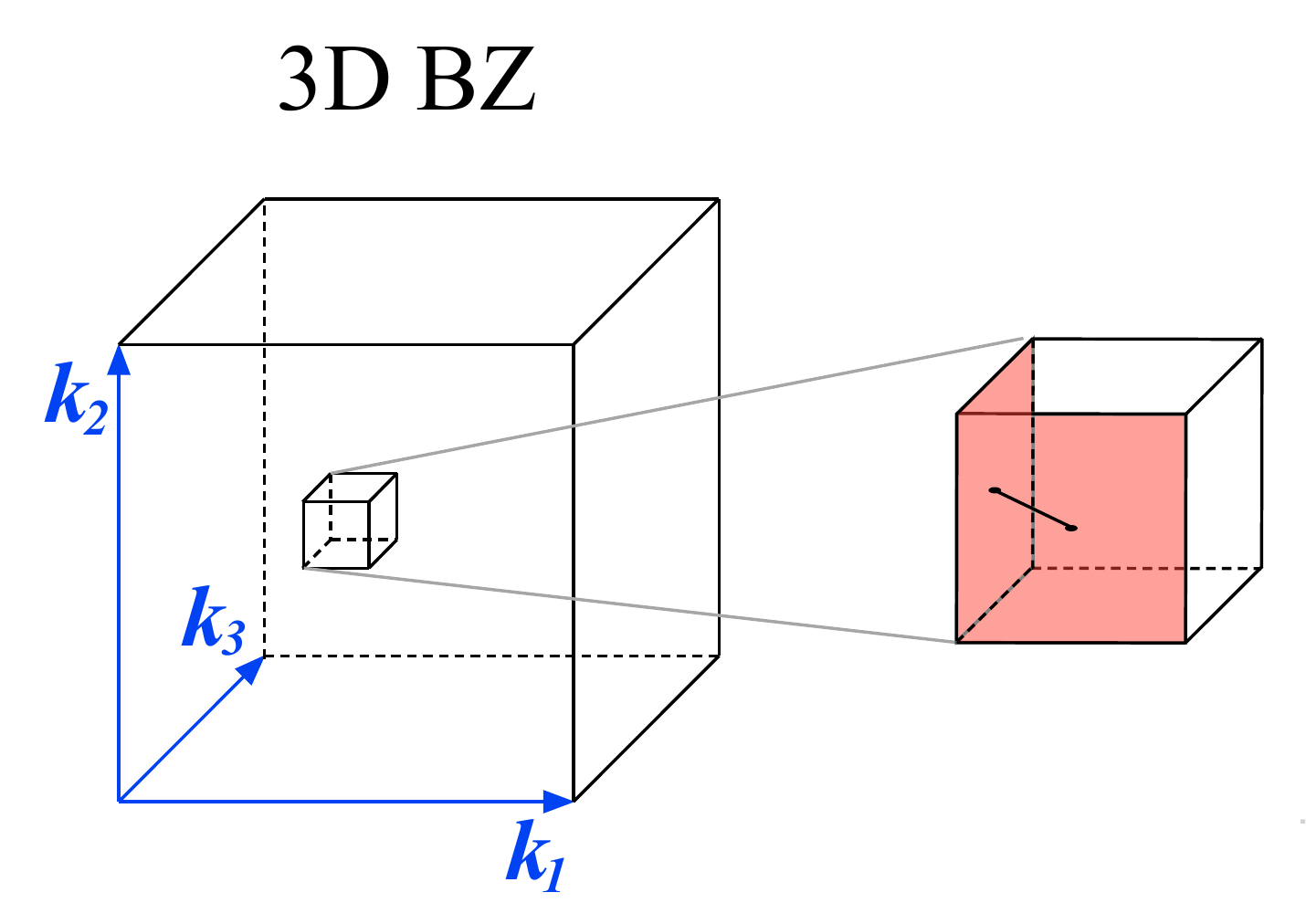}
  \caption{Schematic picture of drawing a line. Red colored surfaces indicate Berry phase over the edge of the surface equals $\pi$. \label{fig:dis_bp}}
 \end{figure}


In our case, the Berry phase
$\tilde{\upsilon}_{\mu\nu}(\bm{k}_{\bm{l}})$ is quantized into $0$ or
$\pi$ (modulo $2\pi$), owing to the chiral symmetry. Furthermore, 
$\tilde{\upsilon}_{\mu\nu}(\bm{k}_{\bm{l}})=\pi$ implies that there
exist odd number of Dirac points on the associated square
\cite{Kariyado:2015aa,PhysRevB.76.045302,doi:10.7566/JPSJ.82.034712}. In
the 3D view point, a Dirac cone on a square patch means that a line
node threads through that patch. Now, if we set $N_B$ large enough so
that the
possibility of multiple line nodes threading one square patch is
excluded, we can follow each line node by following Berry phase
$\pi$. Namely, as far as the square patches are sufficiently small, we
can draw a reasonably smooth line node by connecting the central points
of the nearby patches with Berry phase $\pi$. Note that we can always
find a nearby patch with Berry phase
$\pi$ when we have a certain patch with Berry phase $\pi$, since a line
node cannot be terminated abruptly. 

\begin{figure}[htbp]
 \begin{minipage}{.33\hsize}
 \centering
    \includegraphics[clip,scale=0.25]{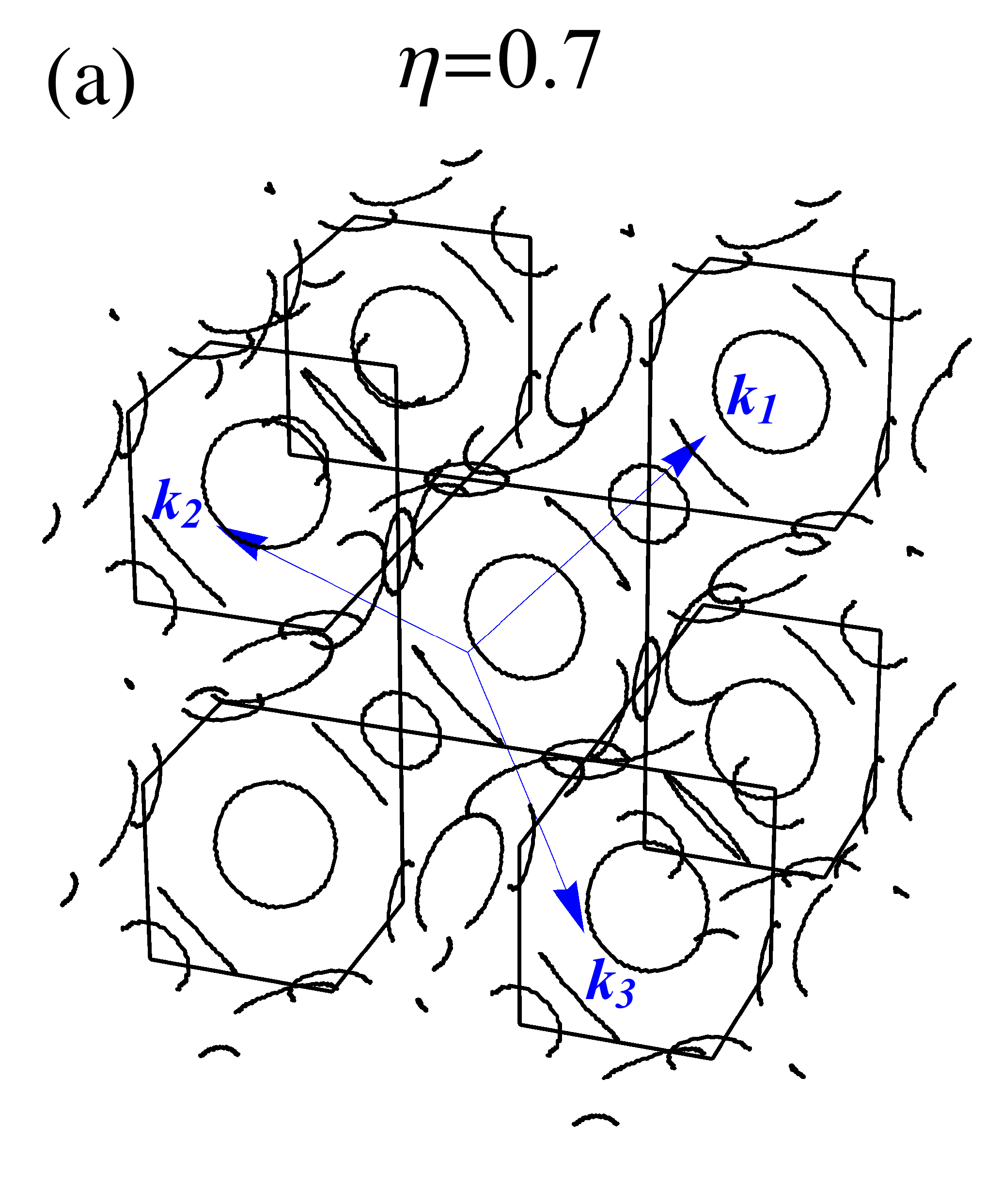}
    \end{minipage}
 \begin{minipage}{.33\hsize}
  \centering
   \includegraphics[clip,scale=0.25]{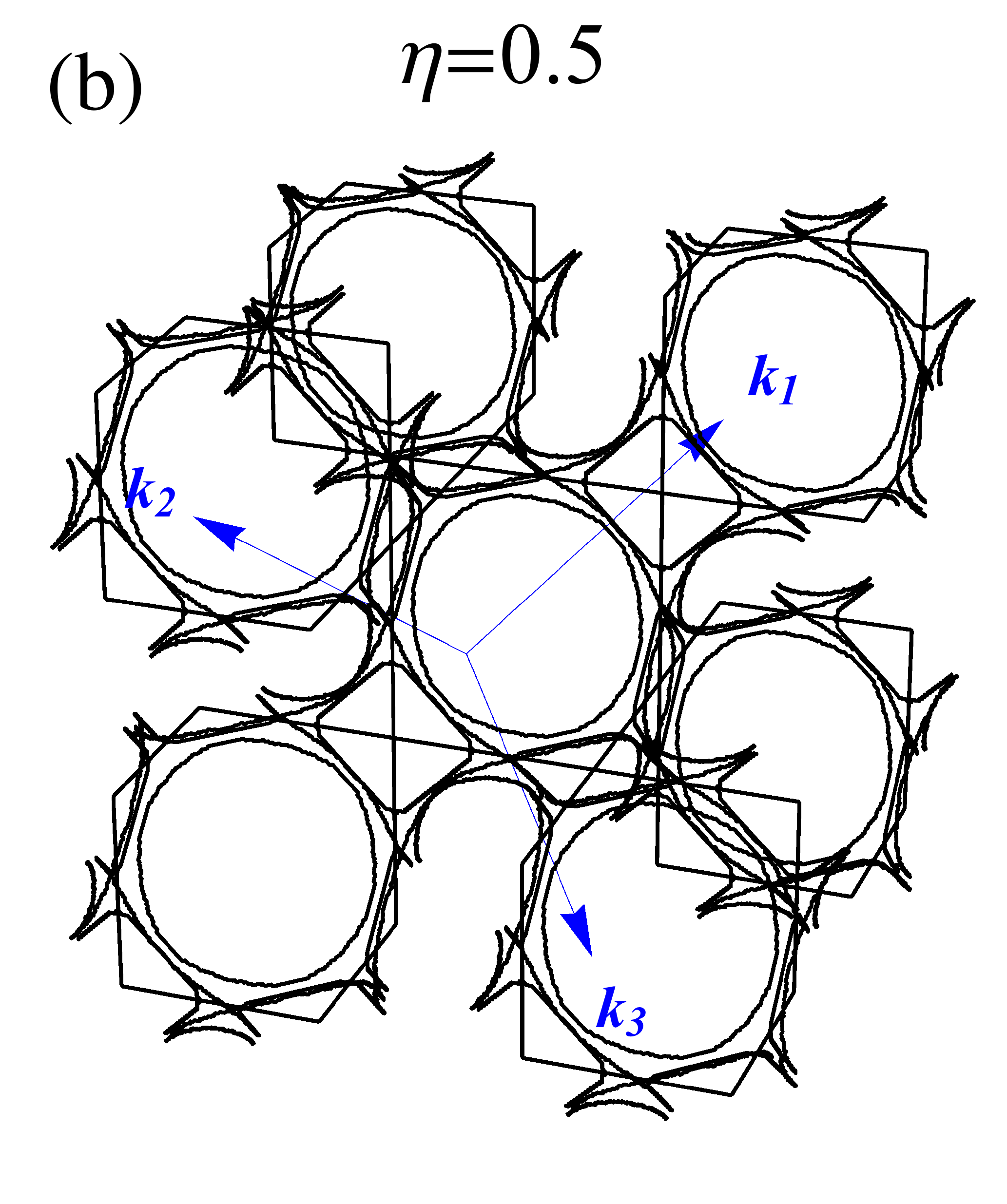}
 \end{minipage}
 \begin{minipage}{.33\hsize}
 \centering
    \includegraphics[clip,scale=0.25]{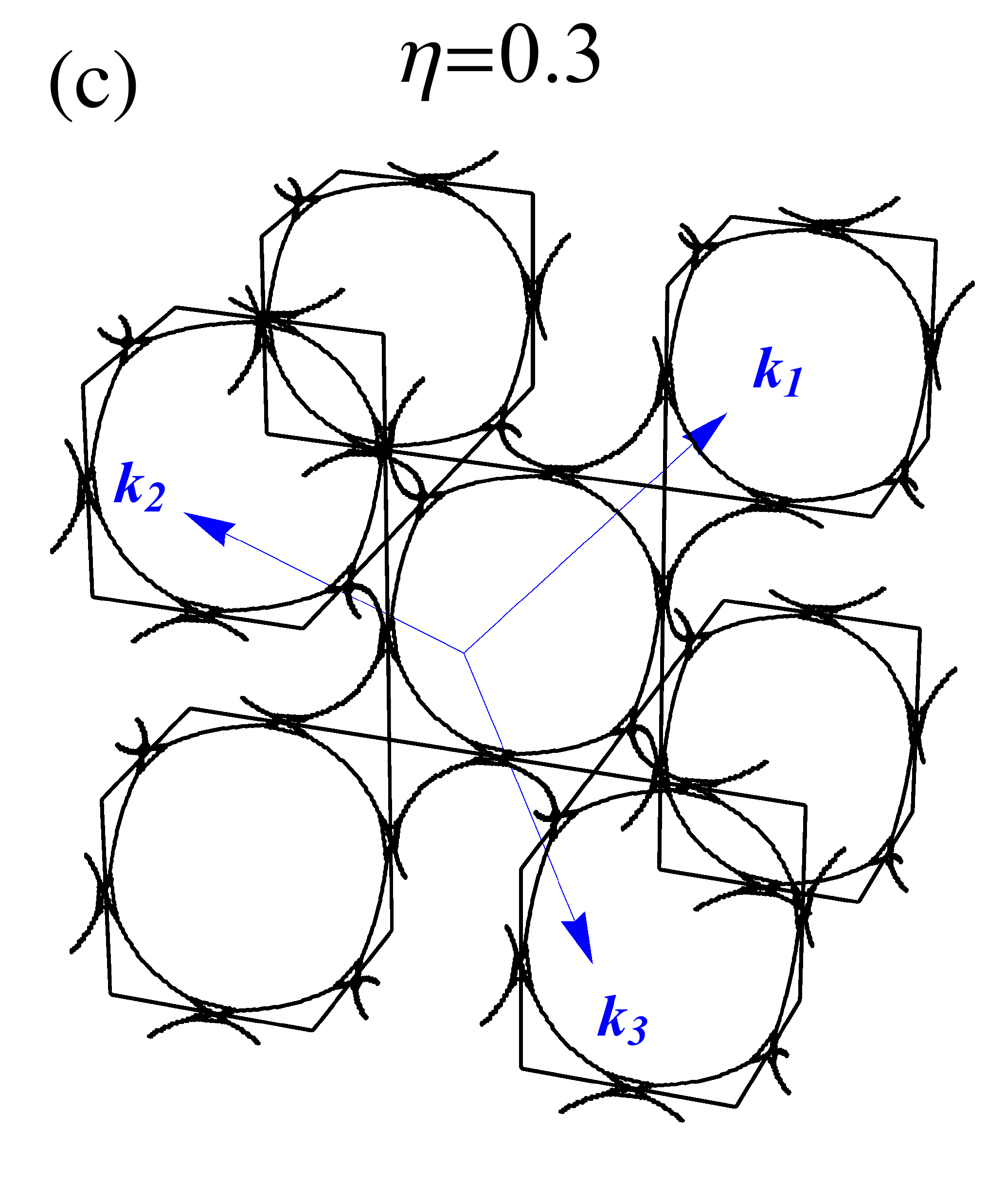}
    \end{minipage}
    \caption{Line nodes for (a) $\eta=0.7$, (b) $\eta=0.5$ and (c) $\eta=0.3$.}
    \label{fig:line}
\end{figure}

Figure \ref{fig:line} illustrates the mapping of the line nodes in the
3D Brillouin zone obtained with $N_B=150$ and $N_D=5$. There are two
classes of the line nodes. The line nodes in the first class are
existing for any $\eta$, and their shapes are 
fixed and identical to those in the single orbital tight-binding model
on the diamond lattice. On the other hand, the line nodes in the second
class appear when $\eta$ becomes smaller than a critical value
$\eta_c=3/4$. The shapes of the line nodes in the second class depend on
$\eta$, and they are eventually absorbed to the line nodes in the first
class at the $\eta=0$ limit.

Before moving onto the next topic, we explain why the gapless points in
our model form a line, 1D object. Although our model is a six band
model, here we focus on a two band effective model that is valid in the
vicinity of the gap closing point. The most generic form of the
effective Hamiltonian can be written as 
\begin{equation}
H(\bm{k})=
\begin{pmatrix}
 R_3 & R_1+\mathrm{i}R_2\\
 R_1-\mathrm{i}R_2 & R_3
\end{pmatrix}
=\bm{R}(\bm{k}) \cdot \bm{\sigma},
\end{equation}
where $\bm{R}(\bm{k})$ is a three-dimensional vector consists of real
quantities $R_1$, $R_2$ and $R_3$, and $\bm{\sigma}$ is a vector with
the Pauli matrices as the components. In our case, the chiral symmetry
induces a constraint on $\bm{R}(\bm{k})$. For simplicity, we choose
$\sigma_3$ as a chiral operator, or assume $\{ H, \sigma_3 \}=0$, which
restricts the hamiltonian to the form
$H(\bm{k})=R_1(\bm{k})\sigma_1+R_2(\bm{k})\sigma_2$. The condition to
have a gap closing point is $|\bm{R}(\bm{k})|=0$, which in fact gives
two conditions $R_1(\bm{k})=0$ and $R_2(\bm{k})=0$. That is, we have two
conditions and three parameters, $k_1$, $k_2$, and $k_3$. Therefore, a
manifold satisfying the zero gap condition has a dimension $3-2=1$,
which supports the existence of the line nodes \cite{Blout1985}.
Note that the choice of $\sigma_3$ as a chiral operator does not mean loss of
generality since it is merely a matter of the basis choice. 

\section{Zak phase and topological edge modes}\label{sec:topological}

\begin{figure}[tbp]
 \centering
  \includegraphics[clip,scale=0.24]{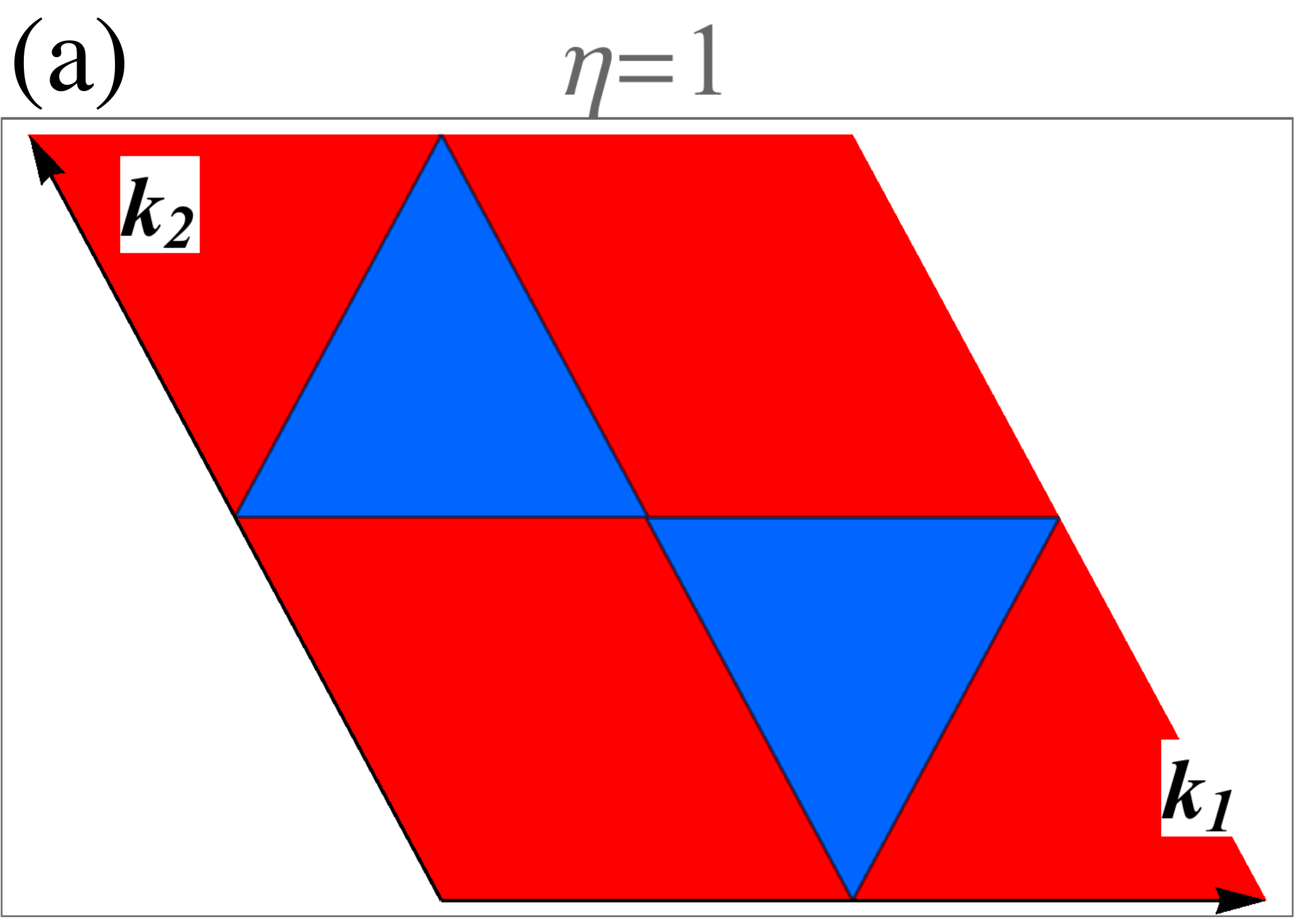}
   \includegraphics[clip,scale=0.24]{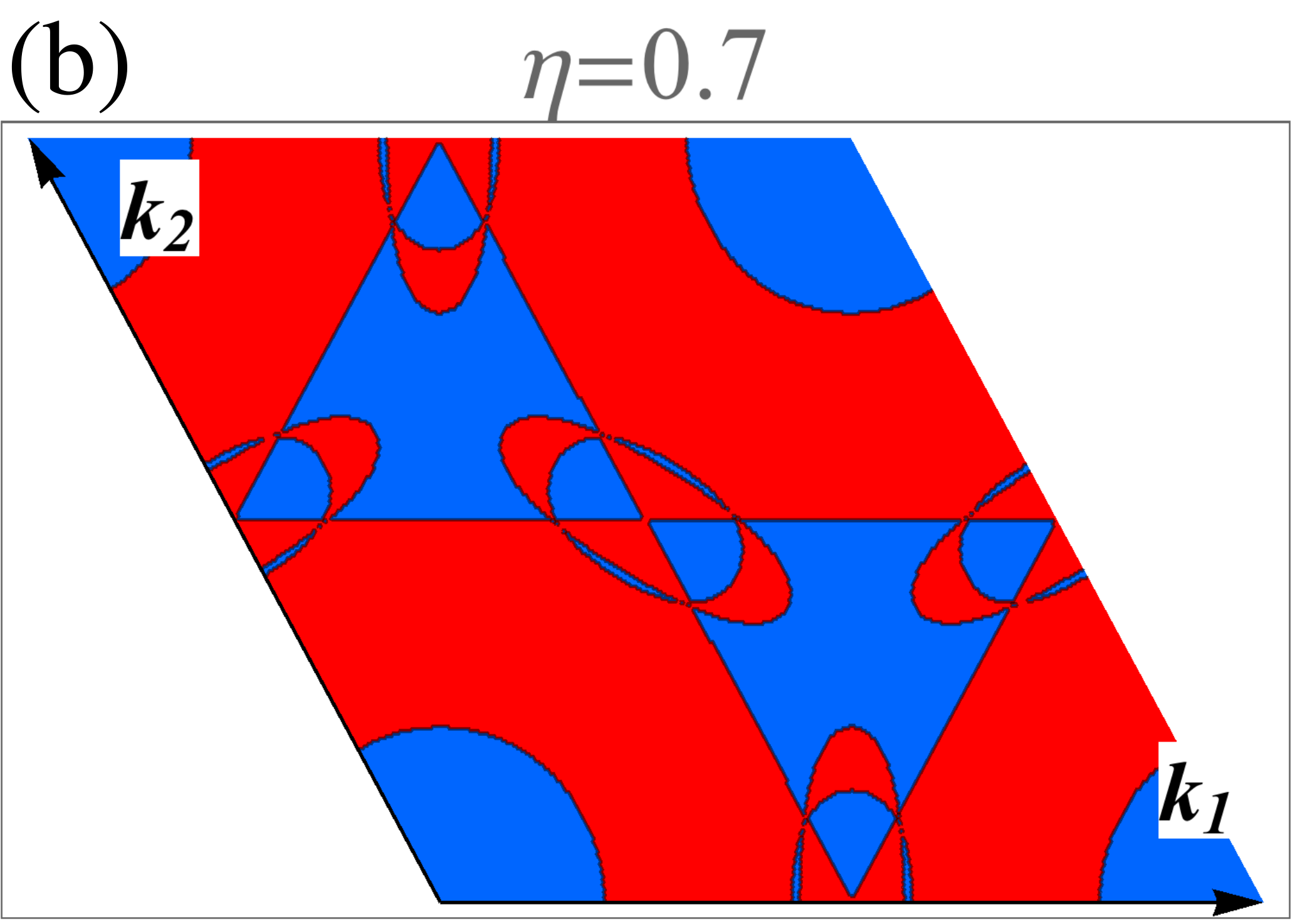}
   \includegraphics[clip,scale=0.24]{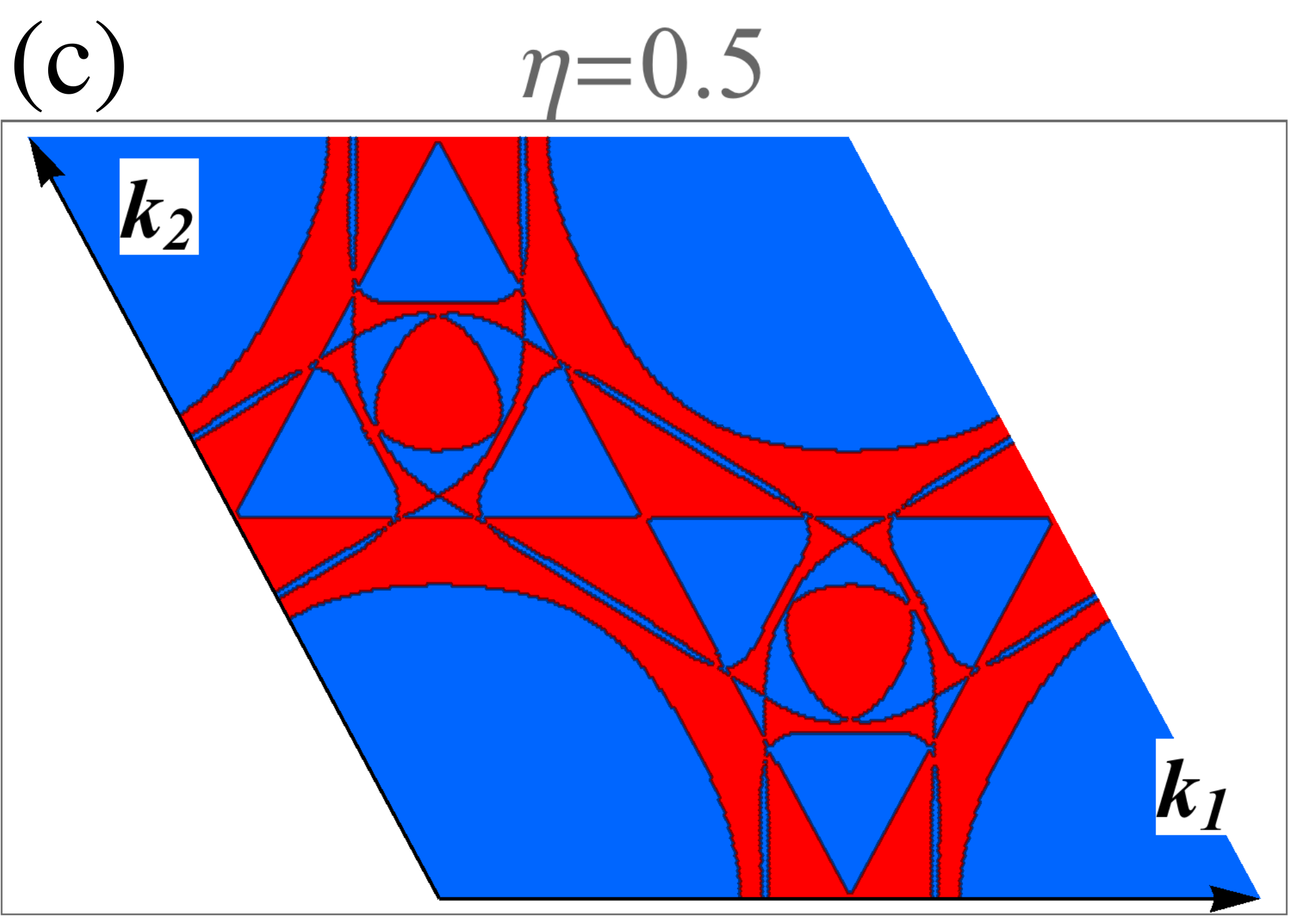}\\
   \includegraphics[clip,scale=0.24]{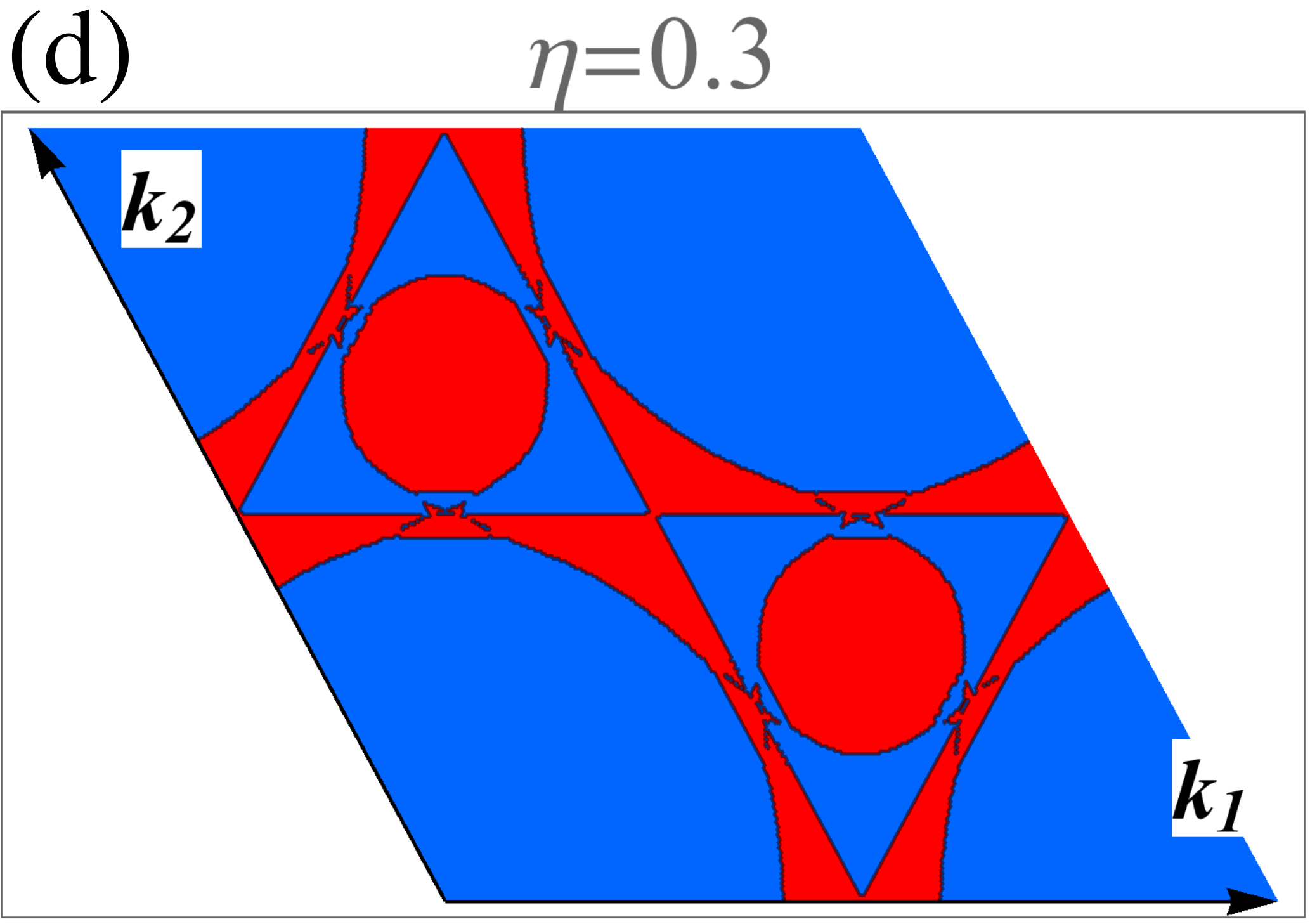}
   \includegraphics[clip,scale=0.24]{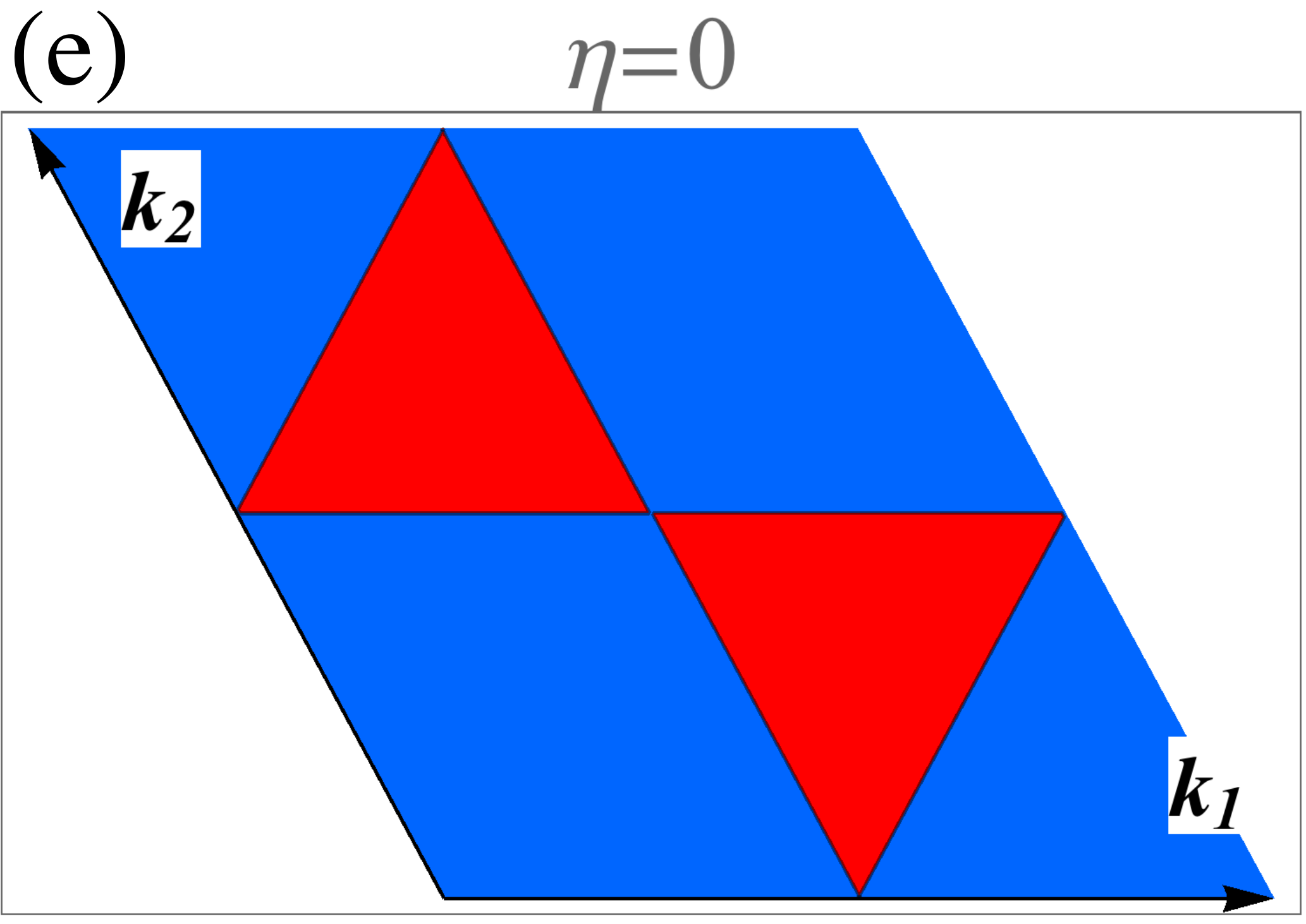}
\caption{Zak phase diagram for (a) $\eta=1$, (b) $\eta=0.7$, (c)
 $\eta=0.5$, (d) $\eta=0.3$ and (e) $\eta=0$.}
\label{fig:zak}
\end{figure}

In order to relate nontrivial topology induced by the bulk band
singularity and edge modes, we calculate the Zak phase, whose definition
is
\begin{equation}
 \upsilon(k_1,k_2)=-\mathrm{i}\int_{L} \mathrm{Tr}\,\mathrm{d}
  \mathcal{A}\Bigr|_{k_1:\mathrm{fixed},\, k_2:\mathrm{fixed}}.
\end{equation}
Here, $\mathcal{A}$ is a non-Abelian Berry connection
$\mathcal{A}=\psi^{\dagger}\mathrm{d}\psi$, which is a $3{\times}3$
matrix-valued one-form associated with a multiplet
$\psi=(\ket{n_1}, \ket{n_2}, \ket{n_3})$. The integration path $L$ is
taken to be a periodic path of fixed $(k_1,k_2)$ and
$0\leq{k_3}<2\pi$, a line along $k_3$-direction. Just as in the case of
the Berry phase in the previous section, the Zak phase is quantized into
0 or $\pi$ due to the chiral symmetry \cite{Ryu2002}.  Figures
\ref{fig:zak}(a)-\ref{fig:zak}(e) show
the Zak phase as a function of $(k_1,k_2)$, which is obtained using the
same method as in the previous section by modifying the integration
path. Blue regions indicate $\upsilon(k_1,k_2)=0$, while red regions
indicate $\upsilon(k_1,k_2)=\pi$. The obtained results are consistent
with the three-fold symmetry and the reflection symmetry of the diamond
lattice. The lines dividing the blue and red regions are the line nodes
(See Fig.~\ref{fig:line}) projected on to the surface under
consideration, that is, the jumps in $\upsilon(k_1,k_2)$ are associated
with the bulk singularity. The complicated structures for intermediate
$\eta$ can be understood by relating each of the loop on the surface to
each of the line node in the bulk. This point will be discussed later.

\begin{figure}[tbp]
 \centering
   \begin{minipage}{.33\hsize}
   \includegraphics[clip,scale=0.26]{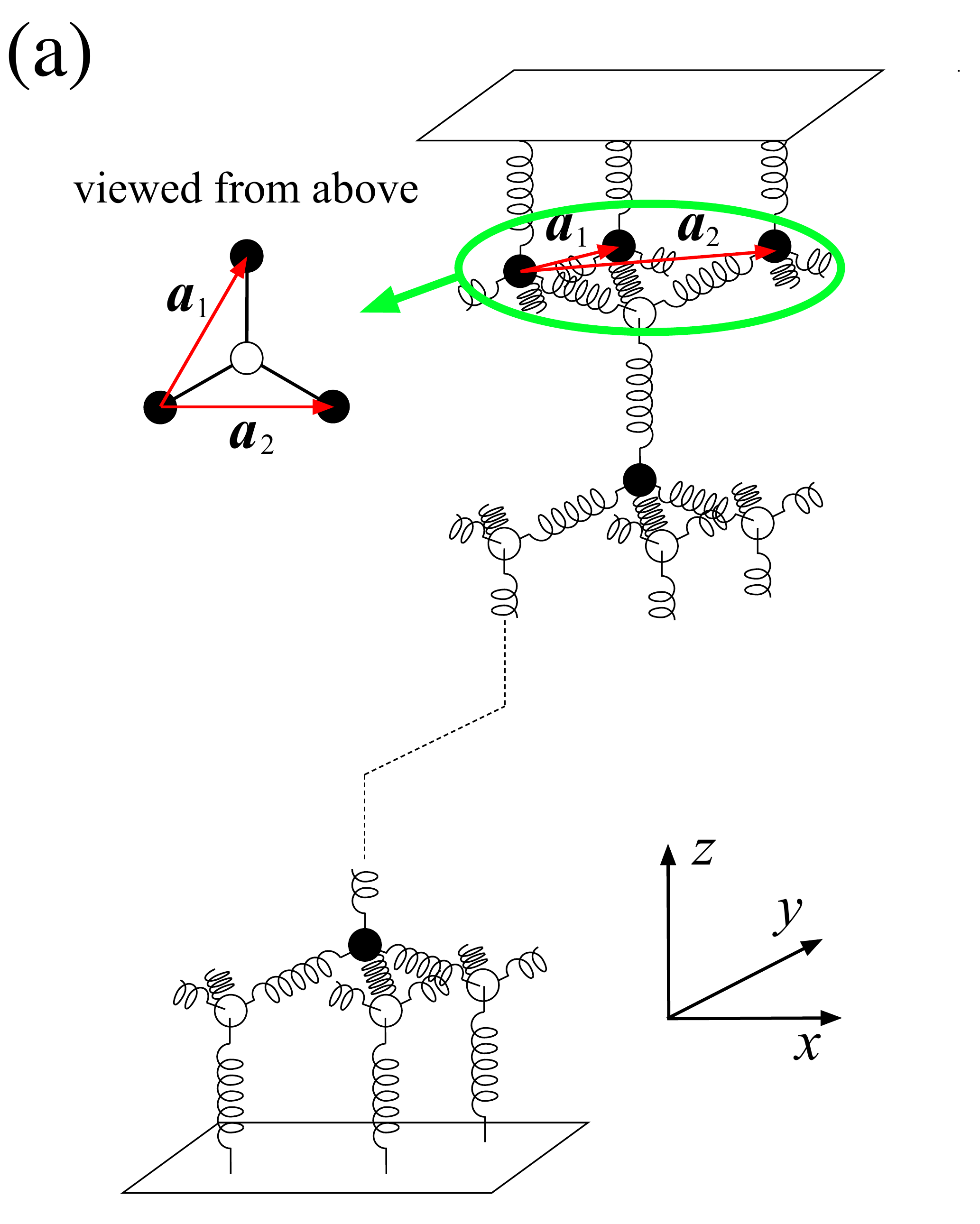}
    \end{minipage}
   \begin{minipage}{.66\hsize}
   \includegraphics[clip,scale=0.24]{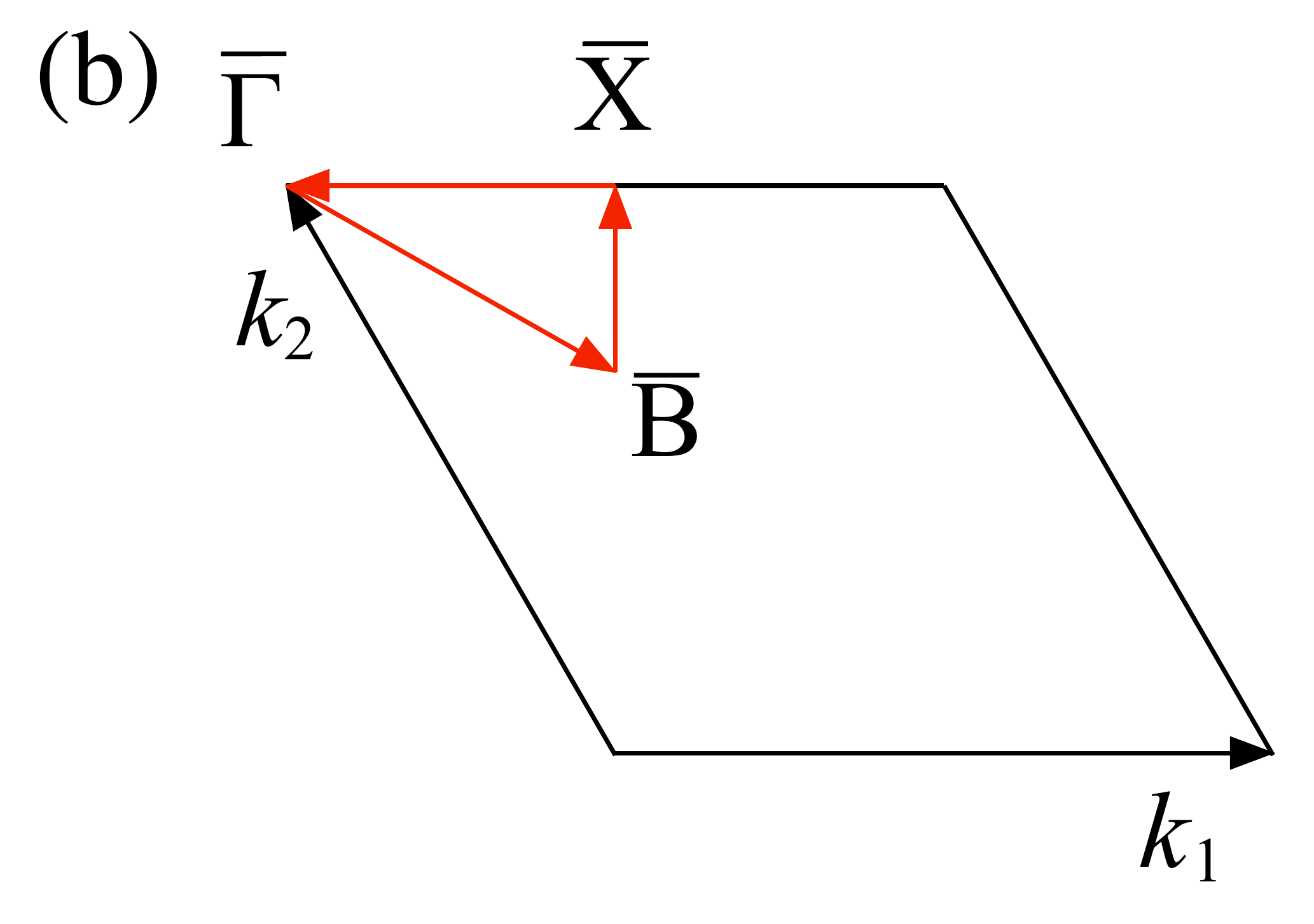}
   \includegraphics[clip,scale=0.24]{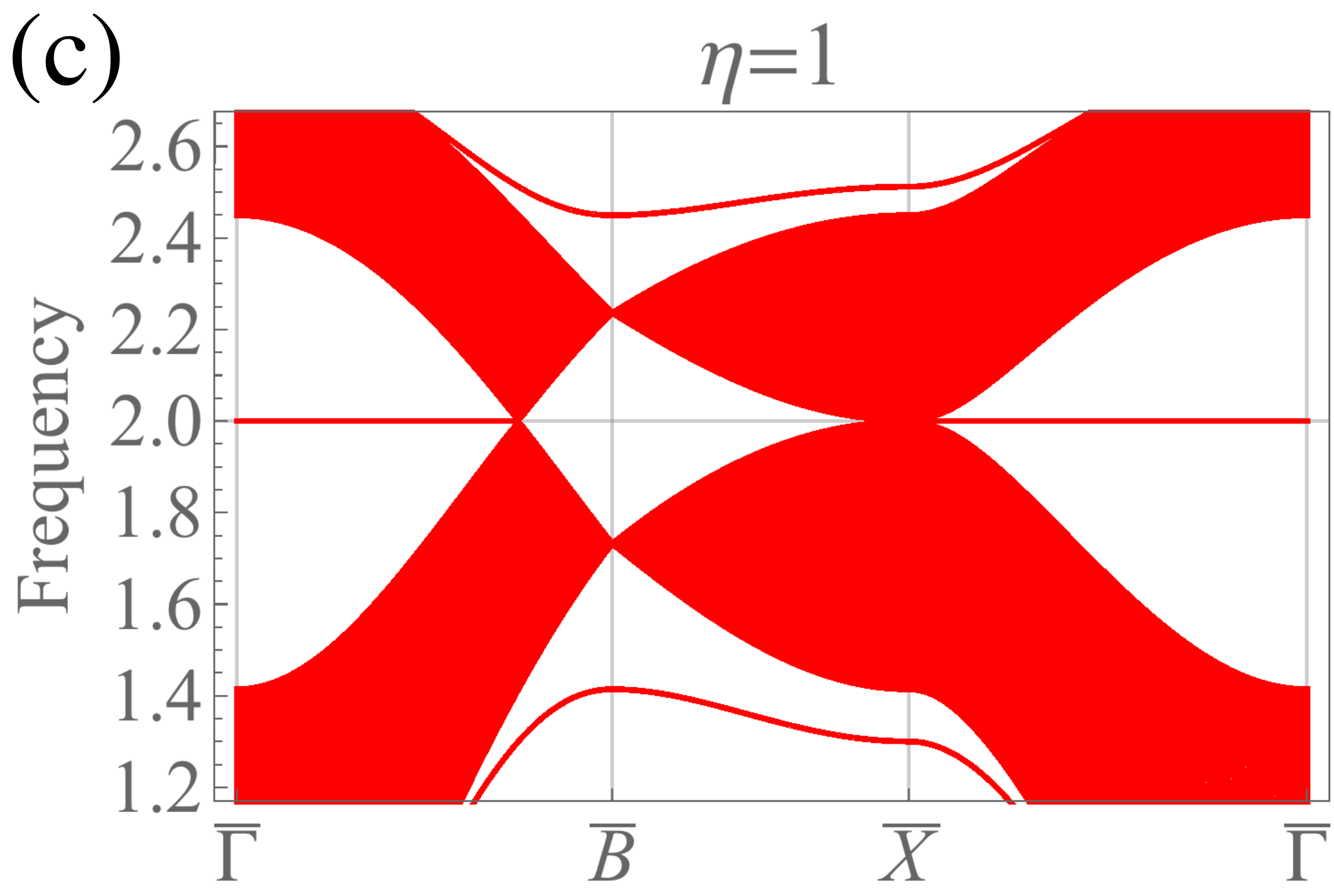}\\
   \includegraphics[clip,scale=0.24]{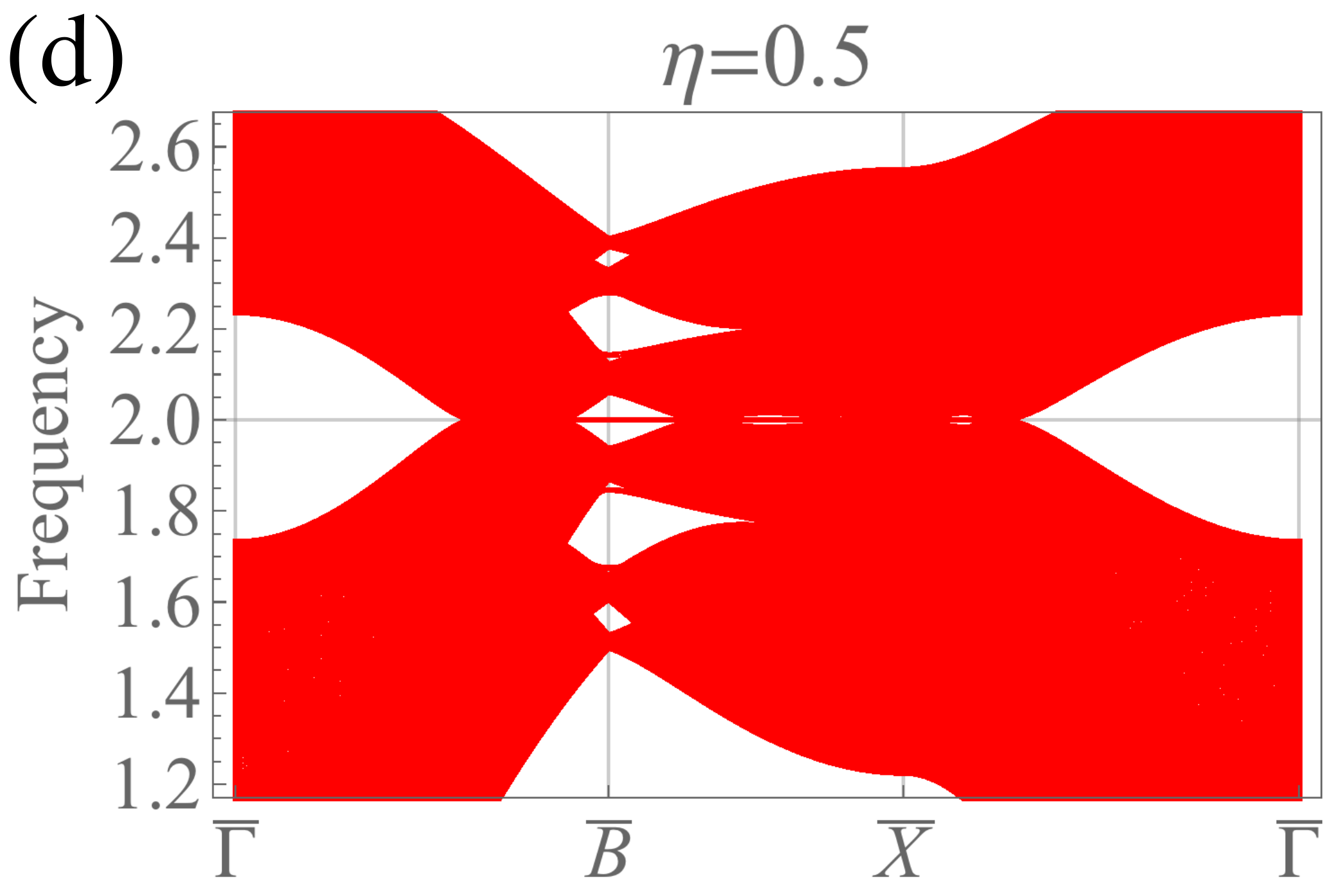}
   \includegraphics[clip,scale=0.24]{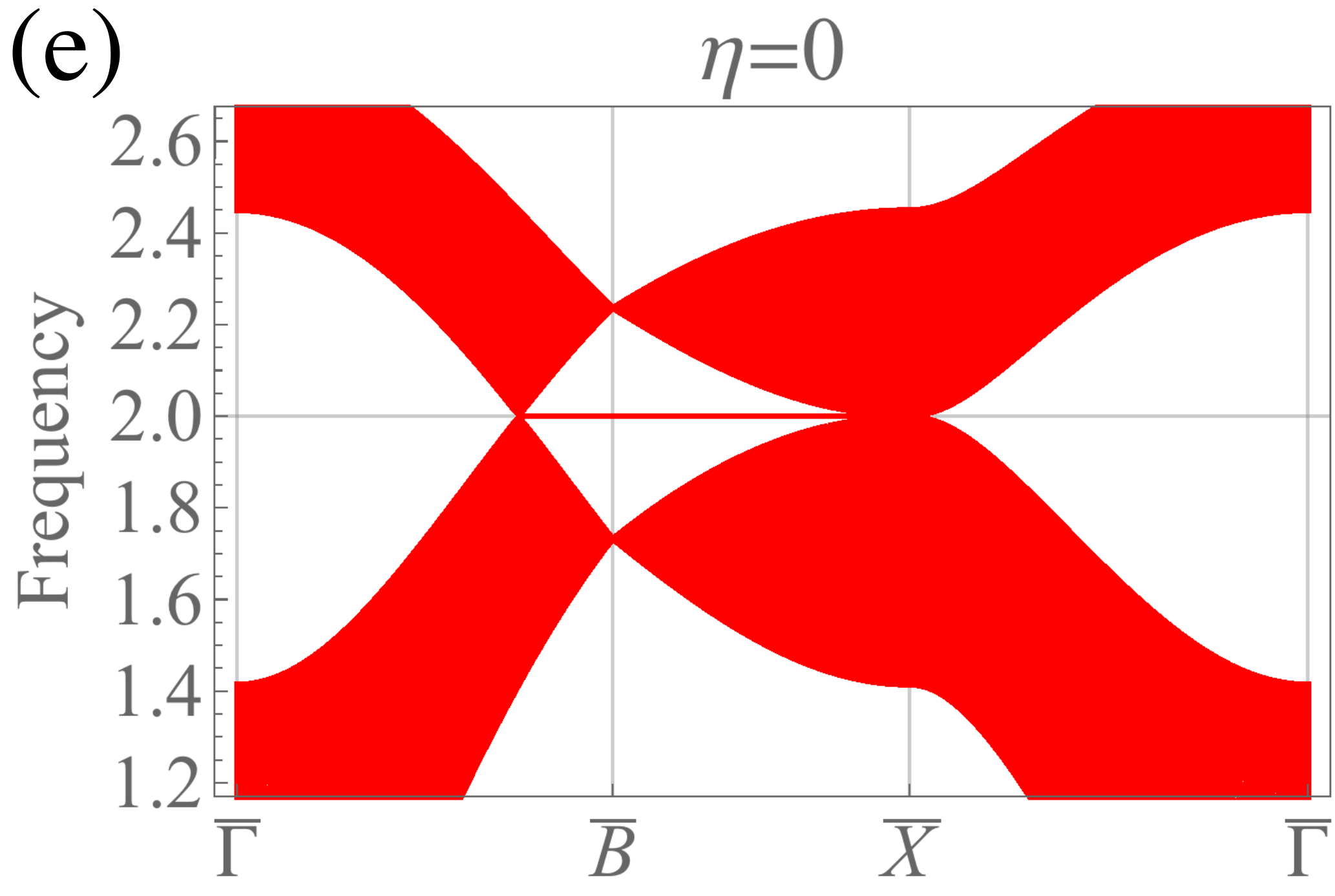}
   \end{minipage}
\caption{
 (a) Schematic picture of the system with surfaces and an illustration of
 the fixed boundary condition. (b) Line on which the frequency
 dispersion is calculated. (c)-(e) Band structure with
 surface for (c) $\eta=1$, (d) $\eta=0.5$, and (e) $\eta=0$.}
\label{fig:edge}
\end{figure}

 If we consider a surface parallel
 to the plane spanned by $\bm{a}_1$ and $\bm{a}_2$, on which $k_1$ and
 $k_2$ are good quantum numbers, we expect at least one
 topologically protected edge mode for $(k_1,k_2)$ with
 $v(k_1,k_2)=\pi$. This statement can be confirmed by investigating edge
 mode explicitly. For that purpose, we consider a system periodic in
 the $\bm{a}_1$ and $\bm{a}_2$ direction, but finite in the $\bm{a}_3$
 direction, so as to have surface parallel to the $\bm{a}_1$--$\bm{a}_2$
 plane as in Fig.~\ref{fig:edge}(a). Here, we apply a fixed boundary
 condition, where the last springs at the surface are connected to the
 wall. This choice of the fixed boundary condition is important to keep
 ``the chiral symmetry'' of the system \cite{Kariyado:2015aa}. Then, the
 frequency dispersion as a function of $(k_1,k_2)$ is obtained by
 diagonalizing $\hat{\Gamma}_{\text{edge}}(\bm{k})$, whose explicit form is
\begin{equation}
 \hat{\Gamma}_{\mathrm{edge}}(\bm k)
  =\begin{pmatrix}
\hat{\mathrm{Z}}    & \hat{\Gamma}_1(\bm k) &                              &                            &           &                              &                            &                             &\\
\hat{\Gamma}^{\dagger}_1(\bm k) & \hat{\mathrm{Z}}   & \hat{\Gamma}_2      &                            &           &                              &                            &                             &\\
                            &  \hat{\Gamma}_2   & \hat{\mathrm{Z}}     & \hat{\Gamma}_1(\bm k) &           &                              &                            &                             &\\
                            &                             & \hat{\Gamma}^{\dagger}_1(\bm k) & \ddots                  & \ddots  &                             &                             &                             &\\
                            &                             &                             &  \ddots                  &           & \ddots                    &                            &                             &\\
                            &                             &                             &                             & \ddots & \ddots                    & \hat{\Gamma}_1(\bm k) &                             &\\ 
                            &                             &                             &                             &           & \hat{\Gamma}^{\dagger}_1(\bm k) & \hat{\mathrm{Z}}    & \hat{\Gamma}_2     &\\      
                            &                             &                             &                             &           &                              & \hat{\Gamma}_2    & \hat{\mathrm{Z}}    & \hat{\Gamma}_1(\bm k) \\ 
                            &                             &                             &                             &           &                              &                            & \hat{\Gamma}^{\dagger}_1(\bm k) & \hat{\mathrm{Z}}        
   \end{pmatrix},
   \label{eq:Gamma_1}
\end{equation}
where $\hat{\Gamma}_1(\bm k)=-\kappa (\hat{\gamma}_4 + \mathrm{e}^{-\mathrm{i} \bm k 
\cdot \bm a_1} \hat{\gamma}_1+\mathrm{e}^{-\mathrm{i} \bm k \cdot \bm a_2}\hat{\gamma}_2)$,
 $\hat{\mathrm{Z}}=\kappa(\hat{\gamma}_1+\hat{\gamma}_2+\hat{\gamma}_3+\hat{\gamma}_4)$ and
 $\hat{\Gamma}_2=-\kappa\hat{\gamma_3}$. Importantly, due to the fixed
 boundary condition, the diagonal part is uniform, which means that the
 eigenvectors are unaffected by the diagonal terms and we still have
 a chiral symmetry in the same sense as the previous analysis.

 Figures \ref{fig:edge}(b)-\ref{fig:edge}(e) show the frequency
 dispersion on a line in the surface Brillouin zone specified in
 Fig.~\ref{fig:edge}(b), obtained using a system with 300 layers in
 $z$-direction.
 We find flat in-gap modes that are localized to the surface. As $\eta$
 reduces from $1$ to $0$, the location of the flat mode changes associated
 with the changes in the bulk spectrum. 
 Notably, there is only one pair of edge modes around $\bar{\Gamma}$
 point for $\eta=1$, while there are three pairs of edge modes around
 $\bar{B}$ point for $\eta=0$. This means that, if we remove the
 degeneracy originating from the existence of two surfaces (top and
 bottom), i.e., focusing on one of the surfaces, the flat edge mode for
 $\eta=1$ is nondegenerate while the one for $\eta=0$ is three-fold
 degenerate.

\begin{figure}[tbp]
 \centering
 \begin{minipage}{.33\hsize}
  \includegraphics[clip,scale=0.4]{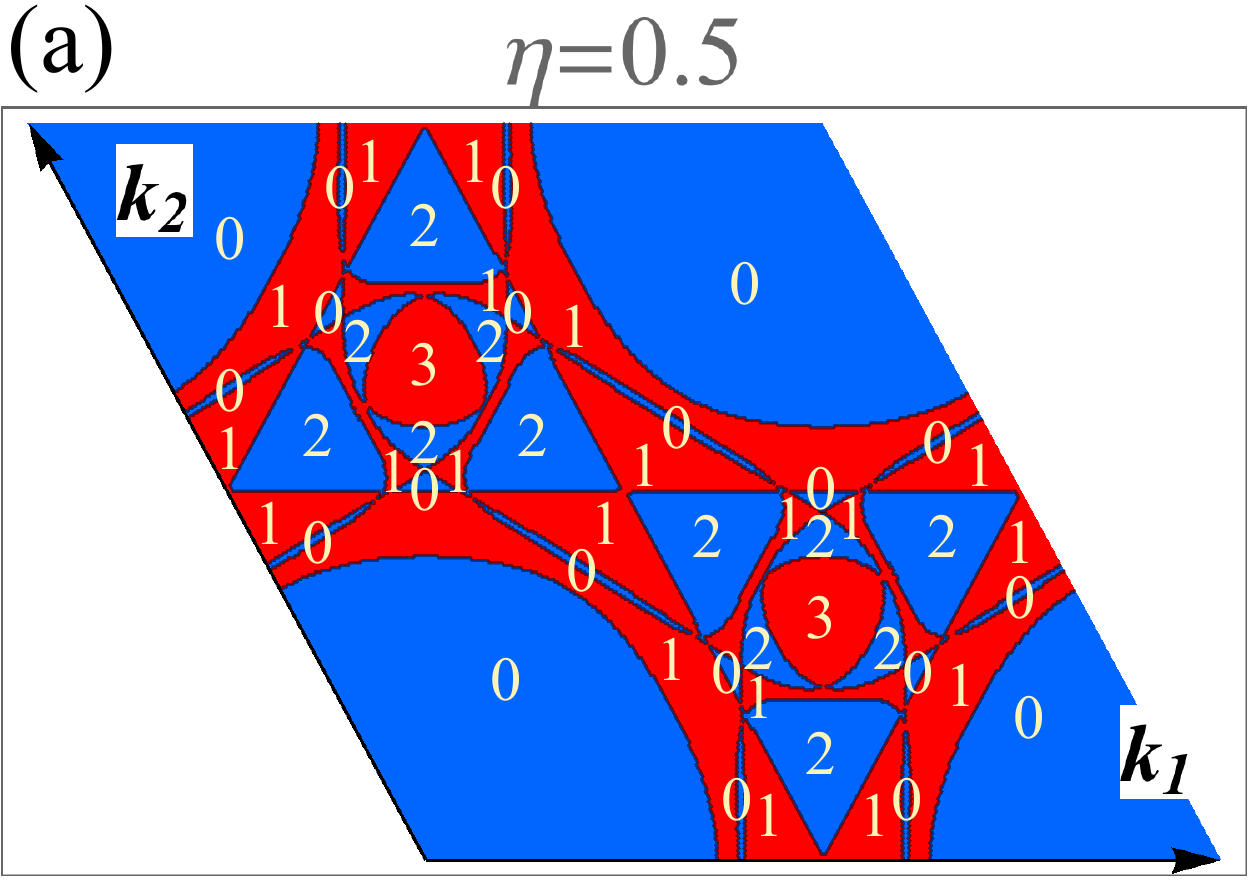}
  \end{minipage}
   \begin{minipage}{.66\hsize}
  \includegraphics[clip,scale=0.35]{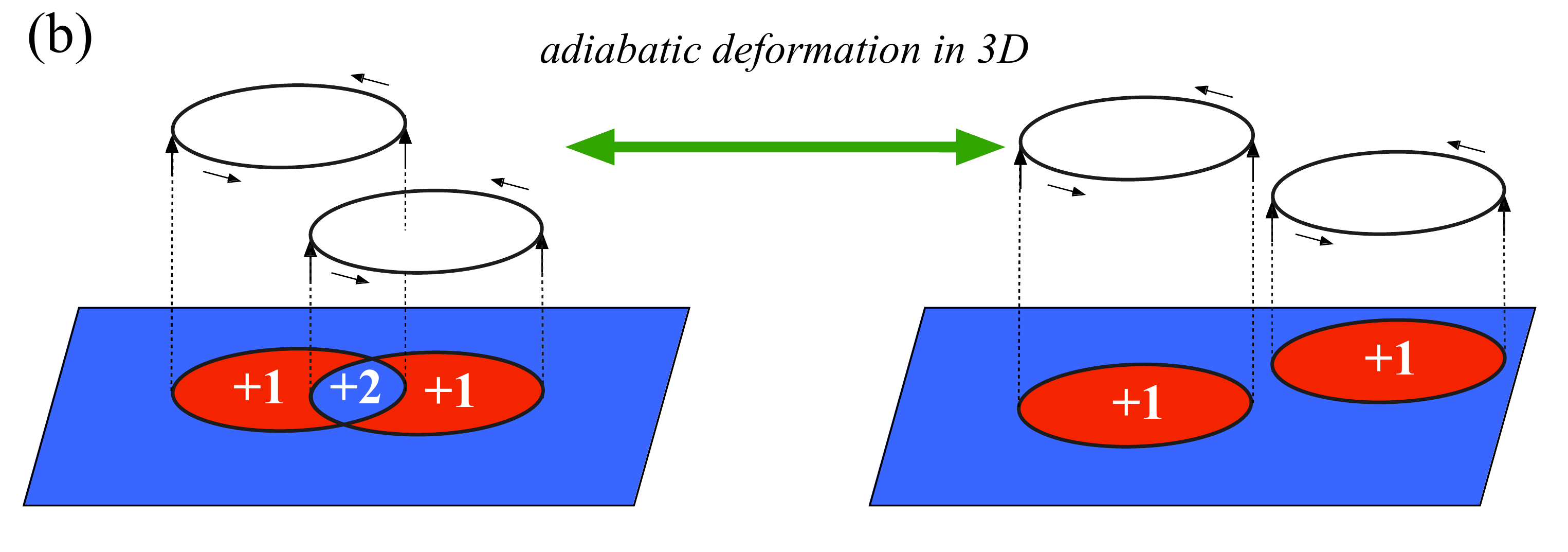}
  \end{minipage}
\caption{(a) Zak phase diagram for $\eta=0.5$ and zero mode edge states. The numbers 
denote the multiplicity of zero modes. (b) Schematic illustration of lifting 2D 
line node loops into 3D Brillouin zone and deforming adiabatically. Arrows 
indicate the chirality of each loops, which all are the same direction.}
\label{fig:zak_zero}
\end{figure}

 Let us relate the edge mode and the Zak
 phase. Figure~\ref{fig:zak_zero}(a) shows the Zak phase and the
 multiplicity of the pair of the edge modes for $\eta=0.5$. As in the
 case of Fig.~\ref{fig:zak}, the blue and red regions correspond
 $\upsilon(k_1,k_2)=0$ and $\pi$, respectively. From this picture, it is
 found that even number of pairs exist for the region with
 $\upsilon(k_1,k_2)=0$, while odd number of pairs for
 $\upsilon(k_1,k_2)=\pi$. It is relatively easy to understand the region
 with multiplicity 0 and 1. For those regions, an established argument
 on the relation between the Zak phase and the edge mode applies as it
 is. Then, the question is how to understand the region with
 multiplicity 2 and 3. Intuitively, it is understood by pulling the
 projected loops in the 2D surface Brillouin zone back into the loops in
 the 3D bulk Brillouin zone. As we can see from
 Fig.~\ref{fig:zak_zero}(b), even when two projected loops are
 overlapping and crossing with each other, there is an
 \textit{adiabatic} way to resolve the overlap in 3D. Assuming that each
 loop carries one edge mode as in the standard argument, and any
 adiabatic change keeps the topological properties intact, the
 multiplicity larger than one can be attributed to the overlapping
 loops.
 
 So far, we have been focusing on the Zak phase so as to emphasize the
 role of the band singularity as a source of the twist in the wave
 function, or eigenvector in our case. However, in order to capture the
 number of the edge modes beyond the parity of the number of the edge
 modes, we have to introduce the other topological invariant, winding
 number. The definition of the winding number is made possible by the
 chiral symmetry. Therefore, the constant diagonal term should be
 subtracted before applying the following argument to our model. For a
 system with the chiral symmetry, an appropriate choice of the basis set
 leads to the hamiltonian of the form
 \begin{equation}
  H=
   \begin{pmatrix}
    0&\hat{\Gamma}\\
    \hat{\Gamma}^\dagger& 0
   \end{pmatrix}.
 \end{equation}
 Then, the winding number is evaluated as \cite{HatsugaiMaruyama2011}
 \begin{equation}
  N_w=-\frac{1}{2\pi}\int_L\mathrm{d}\mathrm{Arg}\,\det\hat{\Gamma},
 \end{equation}
 where the path $L$ is taken to be the same as the path used to define
 the Zak phase that is $(k_1,k_2)$ fixed and $0\leq{k_3}<2\pi$. With
 this choice of the path, the winding number becomes a function of
 $(k_1,k_2)$. Then, it is expected that $N_w(k_1, k_2)$ captures the
 number of edge modes at $(k_1,k_2)$. This expectation is confirmed by
 our numerical calculation of $N_w$. Note that we have a relation
 $\upsilon=\pi N_w$ (mod $2\pi$), indicating that the Zak phase has an
 ability to capture the parity of the number of edge modes. 
 
   Generically speaking, existence of more than one zero modes
   near the same boundary implies gap opening (deviation from zero energy)
   due to the interaction  among  them.
   However in the present model of nearest neighbor spring mass model,
   the edge states of each boundary have fixed chirality, $\chi$,
($\Upsilon\ket{\psi^{\chi}_{\mathrm{edge}}}=
   \chi\ket{\psi^{\chi}_{\mathrm{edge}}}$, $\chi=\pm$) \cite{YH-ch}.
Then due to the selection rule,  $\bra{\psi^{\pm}_{\mathrm{edge}}}(
 \hat{\Gamma}_{\mathrm{edge}}-\kappa(4-\frac{8}{3}\eta)\hat{1})\ket{\psi^{\pm}_{\mathrm{edge}}}=0$, 
 all of the multiple zero mode edge states remain at the zero energy.
 It justifies the winding number itself specifies the number of edge states.
 

\section{Summary}\label{sec:summary}

To summarize, we have investigated the topological properties of
\textit{mechanical diamond}, a three-dimensional spring-mass model with
the diamond structure. We have shown interesting evolution of the
gapless line nodes as a function of $\eta$, which is a parameter
representing tension of the springs in equilibrium. The structure of
line nodes is especially complicated for intermediate $\eta$. In
accordance with the complicated structure of the line nodes, the
multiplicity of the edge modes show the complicated distribution in the
surface Brillouin zone. We have also established the bulk-edge
correspondence in mechanical diamond by relating the edge modes and two
kinds of bulk topological number, the quantized Zak phase and the winding
number, where the chiral symmetry plays an essential role.

\section*{Acknowledgement}
This work is partly supported by Grants-in-Aid for Scientific Research, Nos. 26247064, 25107005, and 16K13845 from JSPS. 

\section*{References}
\bibliographystyle{iopart-num}
\bibliography{diamond}

\end{document}